\documentclass[sigconf]{acmart}

\usepackage{booktabs} 

\usepackage{cancel}
\usepackage{accents}

\usepackage{graphicx}
	\graphicspath{{graphics/}}
	\DeclareGraphicsExtensions{.pdf,.jpeg,.png}
\usepackage[percent]{overpic}
\usepackage{algpseudocode}
\usepackage[caption=false,font=normalsize,labelfont=sf,textfont=sf]{subfig}
\usepackage{enumitem}
\usepackage{tikz}

\setcopyright{rightsretained}

\copyrightyear{2017}
\acmYear{2017}
\setcopyright{acmlicensed}
\acmConference{Middleware '17}{December 11--15, 2017}{Las Vegas, NV, 
USA}\acmPrice{15.00}\acmDOI{10.1145/3135974.3135983}
\acmISBN{978-1-4503-4720-4/17/12}

\begin{document}
\title[SPECTRE: Supporting Consumption Policies in Window-Based Parallel CEP]{SPECTRE: Supporting Consumption Policies in Window-Based Parallel Complex Event Processing}
\author{Ruben Mayer}
\affiliation{%
  \institution{{Institute of Parallel and Distributed Systems }}
  \institution{{University of Stuttgart, Germany}}
}
\email{ruben.mayer@ipvs.uni-stuttgart.de}

\author{Ahmad Slo}
\affiliation{%
  \institution{{Institute of Parallel and Distributed Systems }}
  \institution{{University of Stuttgart, Germany}}
}
\email{ahmad.slo@ipvs.uni-stuttgart.de}

\author{Muhammad Adnan Tariq}
\affiliation{%
  \institution{{Department of Computer Science}}
  \institution{{ FAST - National University of Computer \& Emerging Sciences, Islamabad, Pakistan}}
}
\email{muhammad.adnan@nu.edu.pk}

\author{Kurt Rothermel}
\affiliation{%
  \institution{{Institute of Parallel and Distributed Systems }}
  \institution{{University of Stuttgart, Germany}}
}
\email{kurt.rothermel@ipvs.uni-stuttgart.de}

\author{Manuel Gr\"aber}
\affiliation{%
  \institution{{Institute of Parallel and Distributed Systems }}
  \institution{{University of Stuttgart, Germany}}
}
\email{graeber.manuel@gmx.de}

\author{Umakishore Ramachandran}
\affiliation{%
  \institution{{College of Computing}}
  \institution{{Georgia  Institute of Technology, USA}}
}
\email{{rama@cc.gatech.edu}}

\renewcommand{\shortauthors}{Ruben Mayer et al.}

\begin{abstract}
Distributed Complex Event Processing (DCEP) is a paradigm to infer the occurrence of complex situations in the surrounding world from basic events like sensor readings. In doing so, DCEP operators detect event patterns on their incoming event streams. To yield high operator throughput, data parallelization frameworks divide the incoming event streams of an operator into overlapping windows that are processed in parallel by a number of operator instances. In doing so, the basic assumption is that the different windows can be processed independently from each other. However, consumption policies enforce that events can only be part of one pattern instance; then, they are consumed, i.e., removed from further pattern detection. That implies that the constituent events of a pattern instance detected in one window are excluded from all other windows as well, which breaks the data parallelism between different windows. In this paper, we tackle this problem by means of speculation: Based on the likelihood of an event's consumption in a window, subsequent windows may speculatively suppress that event. We propose the SPECTRE framework for speculative processing of multiple dependent windows in parallel. Our evaluations show an up to linear scalability of SPECTRE with the number of CPU cores.
\end{abstract}

%
%




\maketitle

\begin{tikzpicture}
\begin{scope}[overlay]
\node[text width=20cm] at ([yshift=-13cm,xshift=-9cm]current page.south) {(c) Owner 2017. This is the authors' version of the work. It is posted here for your personal use. Not for redistribution. \newline The definitive version is published in Proceedings of Middleware '17, Las Vegas, NV, USA, December 11--15, 2017, http://dx.doi.org/10.1145/10.1145/3135974.3135983.};
\end{scope}
\end{tikzpicture}

\section{Introduction}

Distributed Complex Event Processing (DCEP) \cite{Jain:2006:DIE:1142473.1142522, Schultz-Moller:2009:DCE:1619258.1619264} is a paradigm applied in many different application areas like logistics, traffic monitoring, and algorithmic trading, to infer the occurrence of complex situations in the surrounding world from basic events like sensor readings or stock quotes. Such situations can be, for instance, the delayed delivery of a packet, traffic jams or accidents and leading market signals. In order to stepwise infer their occurrence from the sensor streams, a distributed network of interconnected DCEP operators, the \emph{operator graph}, is deployed. Each operator processes incoming event streams and detects a designated part of an event pattern that corresponds to a situation of interest. If such a pattern is detected, a new (complex) event is produced and emitted to successor operators or to a consumer, i.e., an entity interested in the corresponding situation. In doing so, operators face increasingly high event loads from their incoming event streams.

In order to be capable of processing high load, the parallelization of DCEP operators has been proposed. In this regard, data parallelization has proven to be a powerful technique to parallelize operators \cite{hirzelcatalog, Balkesen:2013:RRI:2488222.2488257, 7024105, Mayer:2017:MCO:3093742.3093914, Mayer:2016:GRD:2933267.2933509, Matteis:2017:PPW:3065718.3065760}. Data-parallel DCEP systems split the incoming event streams into independently processable windows that capture the temporal relations between single events posed by the queried event pattern.  The windows are processed in parallel by a number of identical operator instances. An event can be part of different windows, so that windows may overlap.
 
A crucial question in overlapping windows is whether an event can be used in multiple pattern instances or not. In many cases, it is preferable to \emph{consume} an event once it is part of a pattern instance. In particular, this means to not use the same event for the detection of further pattern instances in other windows. This way, semantic ambiguities and inconsistencies in the complex events that are emitted can be resolved or prevented. The problem tackled in this paper is that event consumptions impose dependencies between the different windows and thus, prevent their parallel processing. When the same event is processed in parallel in two different windows, consuming it in the first window also consumes it from the second window; hence, there is a dependency between both windows, which can hinder their parallel processing. Understanding that problem, it is no surprise that existing parallel implementations of DCEP systems \cite{Cugola2012205, Balkesen:2013:RRI:2488222.2488257, 7024105} do not support event consumptions, whereas sequential systems often do \cite{Chakravarthy1994, Adi:2004:ASM:988145.988150, Cugola:2012:CEP:2221990.2222305}. This limits the scalability of operators that impose event consumptions. Moreover, it even impedes event consumptions from their further development in academia and industry, as in times of Big Data and Internet of Things, parallel DCEP systems are becoming the gold standard. 
 

In this paper, we propose a speculative processing method that allows for parallel processing of window-based DCEP operators in case of event consumptions. The basic idea is to speculate in each window which events are consumed in the previous windows---instead of waiting until the previous windows are completely processed. This way, multiple overlapping windows can be processed in parallel despite inter-window dependencies. To this end, we propose the SPECTRE (SPECulaTive Runtime Environment) framework, comprising the following contributions: (1) A speculative processing concept that allows the execution of multiple versions of multiple windows using different event sets in parallel. (2) A probabilistic model to process always those window versions that have the highest probability to be correct. (3) Extensive evaluations that show the scalability with a growing number of CPU cores.

\section{Background and Problem Analysis}
\label{sec:background}

To solve the problem of parallel event processing in face of event consumptions, we first discuss a common DCEP model in Section \ref{sec:dcep}.  In Section \ref{sec:parallelization}, we analyze existing DCEP operator parallelization methods and highlight the properties of \emph{window-based data parallelization} as an expressive and scalable parallelization method \cite{7024105, Mayer:2017:MCO:3093742.3093914, Matteis:2017:PPW:3065718.3065760}. Finally, in Section \ref{sec:challenges} we explain the challenges on parallel processing imposed by event consumptions.

\subsection{DCEP Systems}
\label{sec:dcep}
A DCEP system is modeled as an operator graph which inter-connects event sources, operators and consumers by event streams. An event $e$ consists of attribute-value pairs containing meta-data, such as event \emph{type}, sequence numbers or timestamps, and the event \emph{payload}, such as sensor readings, stock quotes, etc. Based on the event meta-data, events from different streams arriving at an operator have a well-defined global ordering (e.g., by timestamps and tie-breaker rules). Each operator $\omega$ processes events in-order on its incoming streams, detecting event patterns according to a pattern specification. If a pattern instance is detected, the operator emits a (complex) event to its successor in the operator graph.

Event patterns are specified in an \emph{event specification language} such as Snoop \cite{Chakravarthy1994}, Amit \cite{Adi:2004:ASM:988145.988150}, SASE \cite{wu2006high}, or Tesla \cite{Cugola:2010:TFD:1827418.1827427}. Those languages involve operators like event sequences, conjunctions, and negations, in order to define the event patterns to be detected. To express the set of relevant events in pattern detection, the pattern specification imposes a \emph{sliding window} of valid events \cite{Arasu:2006:CCQ:1146461.1146463, Matteis:2017:PPW:3065718.3065760}. This can depend on time or the number of events \cite{wu2006high, Cugola:2010:TFD:1827418.1827427, Cugola:2012:CEP:2221990.2222305}, but also on more complex predicates, e.g., on (combinations of) specific event occurrences that mark the beginning and end of a window \cite{7024105}. In this paper, we denote the valid window at a specific point in time as $w_i$. When the window slides, the subsequent valid windows are denoted as $w_{i+1}$, $w_{i+2}$ etc.
Depending on the sliding semantics, different subsequent windows can \emph{overlap}, i.e., events are part of multiple different windows.

\textbf{Example:} In intra-day stock trading, an operator $\omega$ receives an event stream containing live stock quote changes of stock $A$ and $B$ throughout the trading day. An analyst wants to detect correlations between a change in $A$ and a change in $B$. To this end, he formulates a query in the Tesla \cite{Cugola:2010:TFD:1827418.1827427} event specification language:\vspace{-0.1cm}
 \begin{equation*}[Q_E]
 \begin{aligned}
 \ & \mathtt{define\ Influence(Factor)} \\
 \ & \mathtt{from\  B()\ and } \\
 \ & \mathtt{A()\ within\ 1min\ from\ B }\\
 \ & \mathtt{where\ Factor = B:change\ /\ A:change }
 \end{aligned}
 \end{equation*}
 \vspace{-0.2cm}

This pattern can be detected by opening a window with a scope of 1 minute whenever an A event occurs; when a B event is detected in a window opened by an A event, a complex event can be created. 

\begin{figure}
\subfloat[CP: None.]{
	\label{fig:policies_es_nc}
    \includegraphics[width=0.48\linewidth]{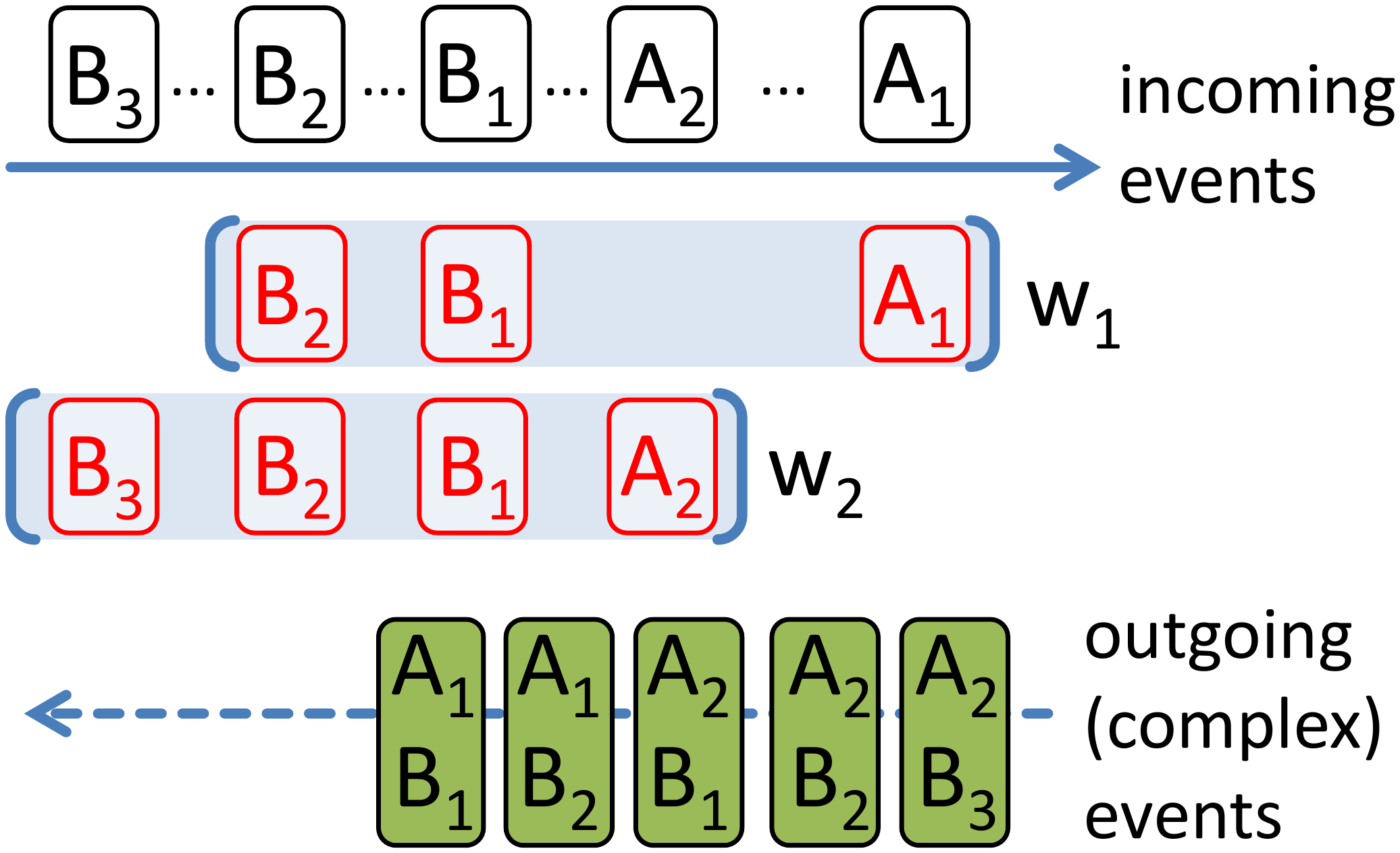}
}
\subfloat[CP: Selected B.]{
	\label{fig:policies_es_sc}
    \includegraphics[width=0.48\linewidth]{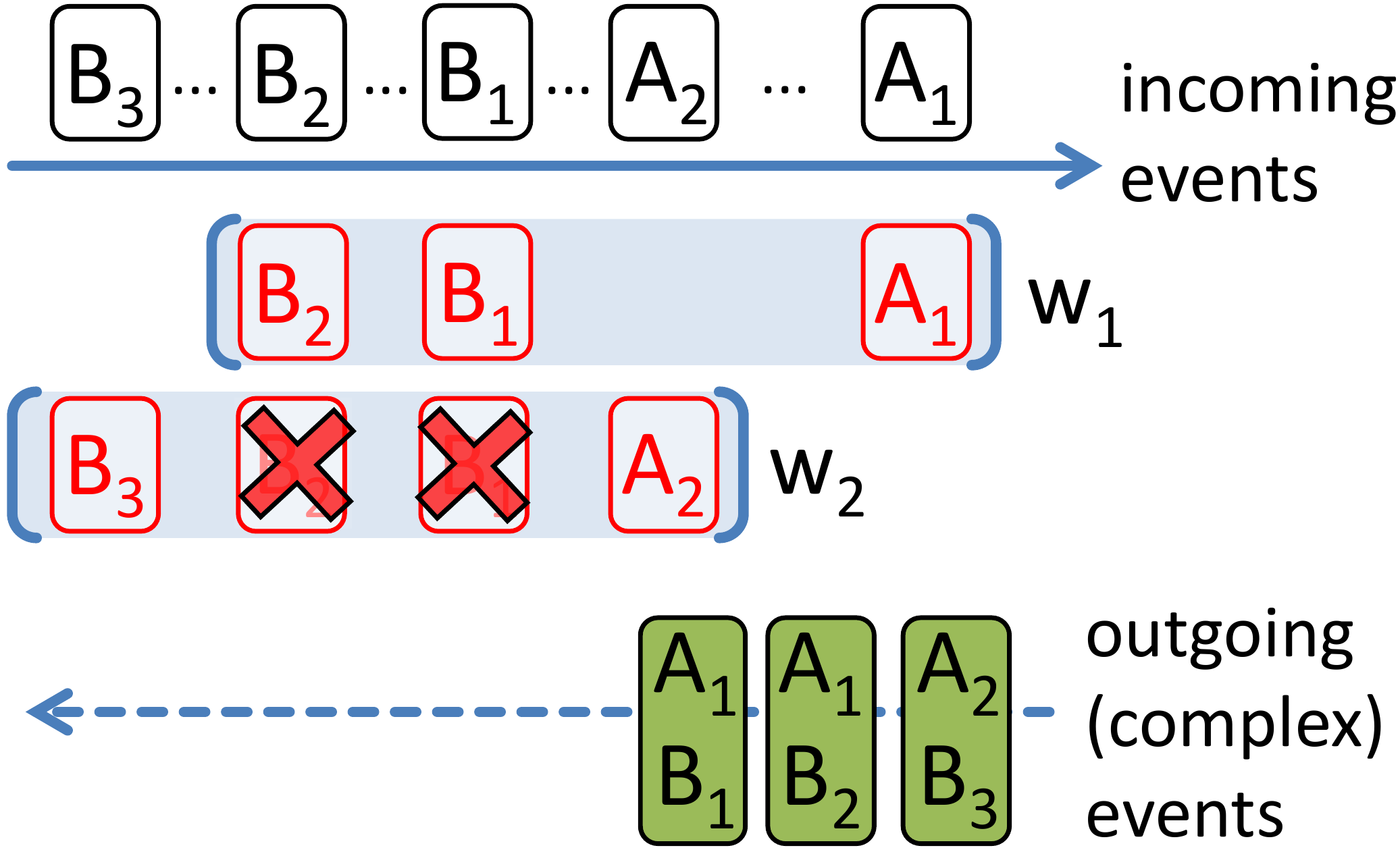}
}
\vspace{-0.2cm}
\caption{Query $Q_E$ with different consumption policies (CP).}
\label{fig:policies}
\vspace{-0.2cm}
\end{figure}

Suppose the events $A_1$, $A_2$, $B_1$, $B_2$ and $B_3$ occur in the event stream in that order, i.e., $A_i$ denotes the $i$-th occurrence of an event of type $A$ in the stream (cf. Figure \ref{fig:policies}). Let us assume that the first $A$ in a window is correlated with every $B$ in the same window---this can be defined in a so-called \emph{selection policy}. As shown in Figure \ref{fig:policies_es_nc}, 5 complex events are detected:\footnote{\ $_{Y}^{X}$ denotes a complex event created from incoming events $X$ and $Y$.} $_{B_1}^{A_1}$, $_{B_2}^{A_1}$, $_{B_1}^{A_2}$, $_{B_2}^{A_2}$, and $_{B_3}^{A_2}$. Notice, that all events are correlated multiple times, i.e., they are not consumed after building a complex event. 

Generally, such multiple correlations of the same event can be problematic. If there is a many-to-one relation between incoming events and detected situations, i.e., many events build a pattern instance but a single event can only be part of one pattern instance, contradicting complex events are produced when events are not consumed. Many-to-one or one-to-one relations are a common case in situation detections.  



Therefore, many event specification languages allow for the specification of a \emph{consumption policy} \cite{Chakravarthy1994, 754955, Adi:2004:ASM:988145.988150, Cugola:2010:TFD:1827418.1827427}. The consumption policy defines which selected events are consumed after they have participated in a complex event detection: It might be \emph{none}, \emph{all} or \emph{some} of them---e.g., depending on the event type or other parameters. A detailed discussion on consumption policies supported in event specification languages is provided in Section \ref{sec:Related Work}. In the example in Figure \ref{fig:policies_es_sc}, selected events of type B are consumed when a complex event is detected, referred to as consumption policy \emph{``selected B''}. Now, only 3 complex events are produced: $_{B_1}^{A_1}$, $_{B_2}^{A_1}$, and $_{B_3}^{A_2}$. In that case, $B_1$ and $B_2$ are not re-used after being correlated with $A_1$ in the first window \textsf{w$_1$}. 


 When a complex event is detected, all constituent events of the event pattern are checked against the consumption policy. Then, all events defined by the consumption policy are consumed as a whole.
This implies that events are not consumed while they only build a \emph{partial match}, but only when the match is completed and a complex event is produced. This inherent property is independent of the concrete selection and consumption policy.
 

\subsection{Operator Parallelization}
\label{sec:parallelization}

The paradigm of \emph{data parallelization} is very powerful in increasing operator throughput. The incoming event stream is split and processed by an elastic number of identical copies of the operator---called \emph{operator instances}. This paradigm has been applied to a wide range of parallel CEP and stream processing systems \cite{7024105, Balkesen2011, zeitler2011massive,  Hirzel:2012:PCP:2335484.2335506, CastroFernandez:2013:ISO:2463676.2465282, Gedik2014, Koliousis:2016:SWH:2882903.2882906, Mayer:2017:MCO:3093742.3093914, Mayer:2016:GRD:2933267.2933509, Matteis:2017:PPW:3065718.3065760}. We assume a shared memory (multi-core) architecture, where the splitter and operator instances are executed by independent threads running on dedicated CPU cores. We assume that the underlying system can provide $k+1$ threads, so that $1$ thread is pinned to the splitter and $k$ threads are pinned to the operator instances. In the rest of this paper, we do not differentiate between operator instances (i.e., instances of the pattern detection logic) and the threads that execute them---we simply refer to both as operator instances.

As mentioned above, we follow a \emph{window-based} data parallelization approach. The incoming event streams are partitioned into windows that capture (temporal) relations defined in the queried pattern. They can naturally be processed by operator instances, as DCEP operators in their core typically work on a sliding window on the event stream \cite{wu2006high, Cugola:2010:TFD:1827418.1827427, Cugola:2012:CEP:2221990.2222305, Koldehofe:2013:RWC:2488222.2488259, 7024105, Mayer:2017:MCO:3093742.3093914, Matteis:2017:PPW:3065718.3065760}. The windows can be based on time, event count or logical predicates that evaluate whether arbitrary window start and end conditions are fulfilled---a more detailed analysis of window-based data parallelization is provided in \cite{7024105} and in \cite{Matteis:2017:PPW:3065718.3065760}. For instance, for the time-based window definition of example query $Q_E$, a new window is opened on each event of type $A$, whereas an open window is closed after 1 minute based on the events' timestamps. The windows are assigned with increasing window IDs and their boundaries are stored in the shared memory (e.g., ``\textsf{w$_{i}$} from event $X$ to event $Y$''). 

The splitter periodically schedules to each operator instance a specific window for processing. The operator instances can hold local state of the processing in shared memory, e.g., partial pattern matches detected in the assigned window. This allows a specific window to be processed by any operator instance at any time; in particular, the processing of a window can be interrupted for some time and resumed later by the same or a different operator instance.

\subsection{Challenges and Goal}
\label{sec:challenges}

In systems without consumptions, processing of a window cannot impact the events within another window, i.e. in principle each pair of windows can be processed in parallel. However, event consumptions impose a dependency between the windows, restricting parallelism, as we discuss in the following. 

Recall the example in Figure \ref{fig:policies_es_sc}. 
The Selection Policy is ``first A, each B'' and the Consumption Policy is ``selected B''. In the first window \textsf{w$_1$}, $A_1$ and $B_1$ build a complex event $_{B_1}^{A_1}$, such that $B_1$ is consumed; furthermore, $A_1$ and $B_2$ build a complex event $_{B_2}^{A_1}$, such that $B_2$ is consumed. If \textsf{w$_1$} and \textsf{w$_2$} are processed in parallel,  the consumption of $B_1$ and $B_2$ in \textsf{w$_1$} might not be known in \textsf{w$_2$}, so that $B_1$ and $B_2$ are erroneously processed in \textsf{w$_2$}, too, leading to inconsistent results. To prevent anomalies due to concurrent processing, \textsf{w$_2$} can only be processed after the consumptions in \textsf{w$_1$} are known. When the event patterns are more complex than in the given minimal working examples, the dependencies become hard to control. For instance, if the pattern requires 3 rising stock quotes of $B$ in a sequence, the completion of the pattern in \textsf{w$_1$}---and hence, the event consumptions---might be unsure until \textsf{w$_1$} is completely processed. If 2 events of type $B$ with rising quotes have already been detected in \textsf{w$_1$}, the completion of the pattern depends on whether a third $B$ occurs; this might only be known at the end of \textsf{w$_1$}.
 The standard procedure to deal with data dependencies is to wait with processing \textsf{w$_2$} until \textsf{w$_1$} is completely processed and hence, all consumptions in \textsf{w$_1$} are known. This, however, impedes the parallel processing of overlapping windows.


\begin{figure}
	\centering
	\includegraphics[width=0.75\linewidth]{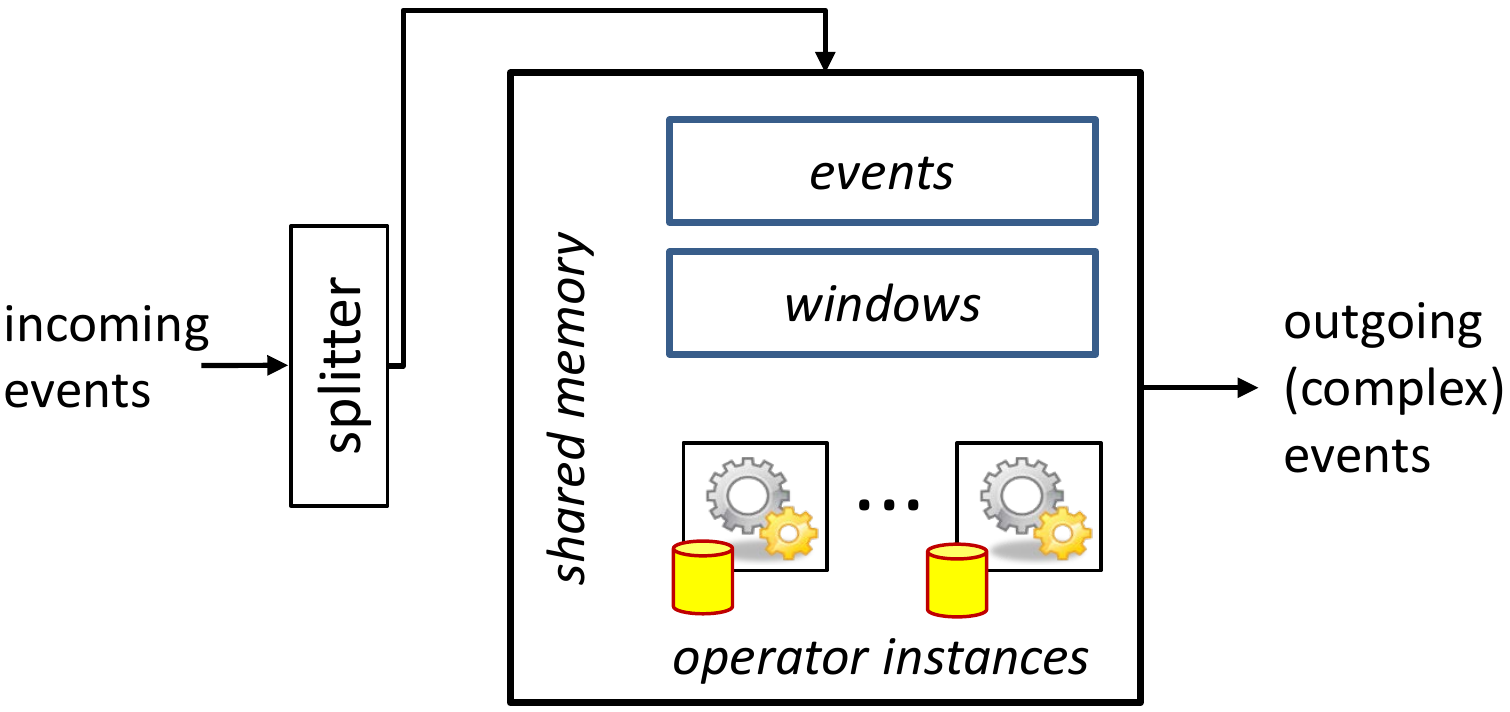}
	\vspace{-0.2cm}
	\caption{Data parallelization framework.}
		\label{fig:parallelization_framework}
\vspace{-0.2cm}
\end{figure}


In this paper, we aim to develop a framework to enable parallel processing of all DCEP operators, regardless of their selection and consumption policy. To this end, we develop a speculative processing method that overcomes the data dependencies imposed by event consumptions, so that data-parallel processing becomes possible. The framework shall deliver exactly those complex events that would be produced in sequential processing; in particular, no false-positive and false-negatives shall occur. 

 \begin{figure*}
	\centering
	\includegraphics[width=0.85\linewidth]{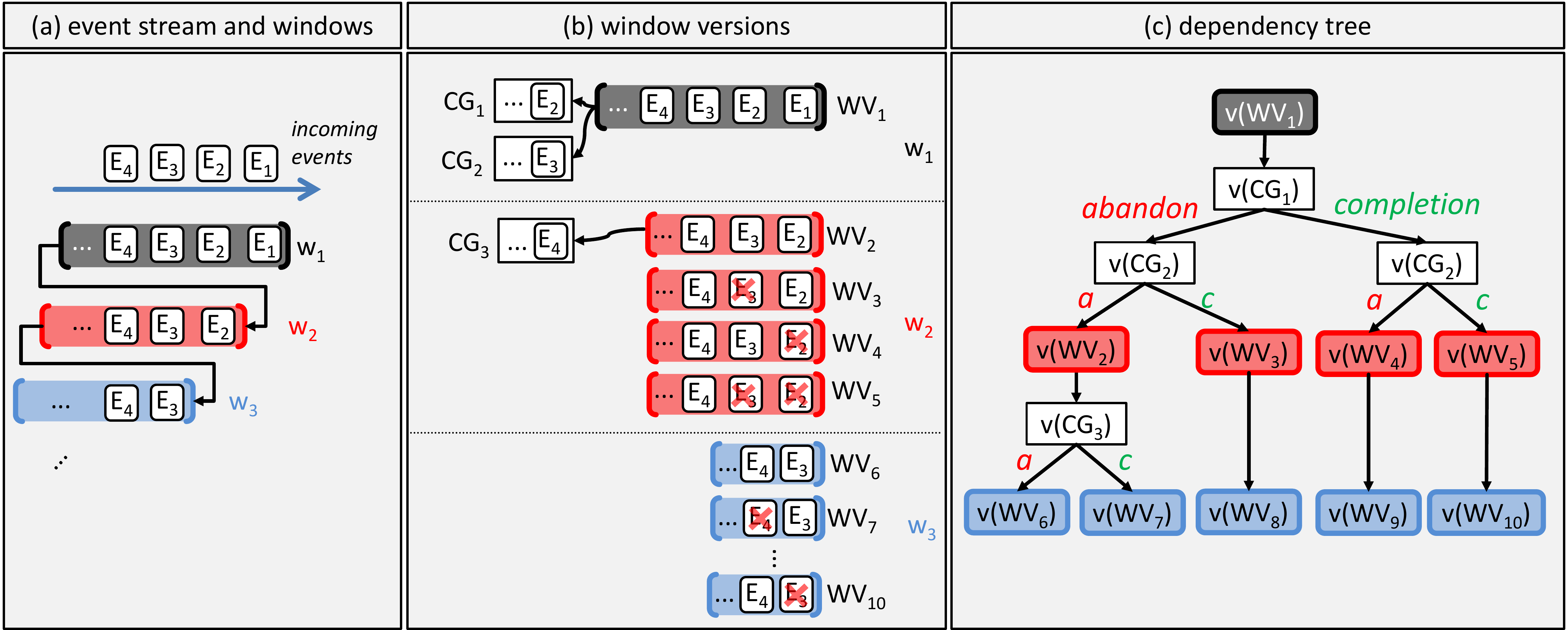}
	\vspace{-0.3cm}
	\caption{Consumption Problem: (a) Structural View. (b) Processing View. (c) Management View.}
\label{fig:version_tree_complete}
\vspace{-0.3cm}
\end{figure*}

\section{The SPECTRE System}

To tackle the dependencies between different windows imposed by event consumptions, we propose the SPECTRE (SPECulaTive Runtime Environment) system, a highly parallel framework for DCEP operators. 
SPECTRE aims to detect the dependencies between different windows and to resolve them by means of speculative execution. 

This section is organized as follows. In Section \ref{sec:speculation}, we introduce the speculative processing approach we follow in SPECTRE. It is based on creating multiple speculative window versions in order to resolve inter-dependencies between windows. Based on that concept, in Section \ref{sec:scheduling}, we explain how SPECTRE determines and schedules the $k$ ``best'' window versions to $k$ operator instances for parallel processing. Finally, in Section \ref{sec:processing}, we provide details on how the  $k$ operator instances perform the parallel processing of the assigned window versions.

\subsection{Speculation Approach}
\label{sec:speculation}

As pointed out above, operators process their incoming data stream based on windows. In particular, operators search for queried patterns to occur in the sequence of events comprised by a window. Windows can overlap, i.e. a pair of windows might have a sequence of events in common. The windows of an operator are totally ordered according to their start events. We call a window, say \textsf{w$_{j}$}, a successor of another window, \textsf{w$_{i}$}, iff the start event of \textsf{w$_{i}$} occurs before the starting event of \textsf{w$_{j}$} in the corresponding event stream. For example, in Figure \ref{fig:version_tree_complete}(a), \textsf{w$_{1}$} starts earlier than \textsf{w$_{2}$}; hence, \textsf{w$_{2}$} is a successor of \textsf{w$_{1}$}. In the same way, \textsf{w$_{3}$} is a successor both of \textsf{w$_{2}$} and of \textsf{w$_{1}$}.

Now, we can define a \emph{consumption dependency} (or \emph{dependency} for short) between windows. Roughly speaking, a window \textsf{w$_{j}$} depends on another window \textsf{w$_{i}$}, if the consumption of some events in \textsf{w$_{i}$} might affect the processing of window \textsf{w$_{j}$}. Formally, we define that \textsf{w$_{j}$} depends on \textsf{w$_{i}$} iff \textsf{w$_{j}$} is a successor of \textsf{w$_{i}$} and \textsf{w$_{j}$} overlaps with \textsf{w$_{i}$}. 
For example, in Figure \ref{fig:version_tree_complete}(a), \textsf{w$_{2}$} depends on \textsf{w$_{1}$}, and \textsf{w$_{3}$} depends both on \textsf{w$_{2}$} and on \textsf{w$_{1}$}.

Now, we will introduce the concept of a \emph{consumption group}. A consumption group is maintained for each partial match of a search pattern found in a window. It records all events of this window that need to be consumed if the partial match becomes a total match, i.e. the corresponding search pattern is eventually detected in the window. Let's assume that an operator is acting on some window \textsf{w}. Whenever the operator processes an event starting a new partial match of some search pattern, it \emph{creates} a new consumption group associated with \textsf{w}. When it processes an event that completes a pattern, it \emph{completes} the corresponding consumption group. On the other hand, a consumption group is \emph{abandoned} if the corresponding pattern cannot be completed anymore. Consequently, while processing the events of a window, multiple consumption groups can be created that are associated with \textsf{w}. However, all of them will be completed or abandoned at the latest when processing of \textsf{w} is finished. 

While acting upon w, the operator adds events to be potentially consumed to the consumption groups associated with w, in conformance with the specified consumption policy. When a consumption group is completed, all events contained in this group are consumed together. If the consumption group is abandoned instead, it is just dropped and no events are consumed.

For example, let us assume that a query for pattern of a sequence of three events of type $A$, $B$ and $C$ in a window of time scope 1 minute, is processed by an operator. Let us further assume the consumption policy is set to consume all participating events in case of a pattern match. When detecting an event of type $A$, say $A_1$, in a window, the operator creates a new consumption group. The first event of type $B$, $B_1$, is added to the consumption group. If the window ends (i.e., 1 minute has passed) and no event of type $C$ is detected, the consumption group is abandoned and no events are consumed in the window. If an event of type $C$, say $C_1$, occurs after $B_1$ and within the window scope, the consumption group is completed, and all three events participating in the pattern match, $A_1$, $B_1$ and $C_1$, are consumed together.

At the time a consumption group is created that is associated with window \textsf{w}, it is unknown whether the corresponding pattern will eventually be completed in \textsf{w}. Clearly, the outcome of the consumption group (complete or abandon) might affect events of all windows that depend on \textsf{w}. One way to handle this uncertainty is to defer the processing of all depending windows until the consumption group terminates (completed or abandoned). However, in general this amounts to processing all windows sequentially. The approach that we follow in SPECTRE is to generate two window versions for each window depending on \textsf{w}, one version assuming that the consumption group will be completed and the other one assuming the consumption group will be abandoned. These window versions can then be processed in parallel to \textsf{w}. Once the outcome of the consumption group is known, i.e., completed or abandoned, processing continues on the corresponding window versions that assume the correct outcome while the other window versions that assume the wrong outcome are just dropped.  Obviously, this approach allows for processing dependent windows in parallel even in the presence of event consumptions.

With this approach, windows that depend on other windows may have multiple versions that depend on the outcome of the associated consumption groups. In principle there is a window version for any combination of the complete and abandon case of the consumption groups that a window depends upon. When one of these consumption group is abandoned, all window versions assuming this consumption group to complete can be dropped, and vice versa.

To capture the dependency between consumption groups and window versions, we introduce the concept of a \emph{dependency tree}. There exists an individual dependency tree for each independent window, i.e., each window that does not depend on any other window according to our definition above. The vertices of the dependency tree are window versions or consumption groups, while the directed edges of the tree specify the dependencies between them. The root of the dependency tree is the only version of an independent window---by definition, there is only one version of an independent window. 

The vertex of a window version $\mathit{WV}$, say v($\mathit{WV}$), has at most one child. The sub-hierarchy rooted by this child includes all versions of windows depending on $\mathit{WV}$, if any. We will denote this sub-hierarchy as v($\mathit{WV}$)'s subtree. The subtree is rooted by a consumption group if a consumption group is associated with v($\mathit{WV}$). Otherwise the root of the subtree is a window version directly dependent on v($\mathit{WV}$), if any.

A vertex representing a consumption group $\mathit{CG}$, say v($\mathit{CG}$), always has two children, one for each possible outcome of $\mathit{CG}$ (completed or abandoned). The so-called completion edge of v($\mathit{CG}$) links the subtree of window versions for which completion of $\mathit{CG}$ is assumed, whereas the so-called abandon edge of v($\mathit{CG})$ links the subtree of window versions which assume $\mathit{CG}$ to be abandoned. That is, all window versions that can be reached via v($\mathit{CG}$)'s completion edge do not include any event included in $\mathit{CG}$, while events in $\mathit{CG}$ have no effect on window versions linked by v($\mathit{CG}$)'s abandon edge.  

When a consumption group $\mathit{CG}$ associated with a window version $\mathit{WV}$ is created, the following is performed: v($\mathit{CG}$) is added as a new child of v($\mathit{WV}$) to the dependency tree. The old subtree of v($\mathit{WV}$) is linked by v($\mathit{CG}$)'s abandon edge, while a modified copy of the subtree is linked by v($\mathit{CG}$)'s completion edge. The modification makes sure that no events included in $\mathit{CG}$ occur in the window versions of the subtree linked by v($\mathit{CG}$)'s completion edge. In other words, for each window version existing in v($\mathit{WV}$)'s old dependent versions subtree, a copy that \emph{suppresses} all events listed in $\mathit{CG}$ is added. Therefore, each new consumption group associated with v($\mathit{WV}$) doubles the window versions in v($\mathit{WV}$)'s subtree.  

\textbf{Examples and Algorithms:}
In the following, a set of examples on the management of the dependency tree is provided along with a formalization of the associated management algorithms. We discuss the following cases: (1) a new dependent window is opened, (2) a new consumption group associated to a window version is created, (3) an existing consumption group is completed or abandoned. 

\emph{New dependent window.} When a new window \textsf{w$_{\mathit{new}}$} is opened that depends on another window \textsf{w$_{x}$}, for every leaf vertex of the dependency tree rooted by the window version of \textsf{w$_{x}$}, new window versions are created as child vertices (Figure \ref{alg:pathtree}, lines 1--10). For example, in Figure \ref{fig:version_tree_complete}, at the start of \textsf{w$_{3}$}, new window versions ($\mathit{WV}_6$ to $\mathit{WV}_{10}$) of \textsf{w$_{3}$} are created and the corresponding vertices ( v($\mathit{WV}_6$) to v($\mathit{WV}_{10}$) ) are attached to all leaf nodes of the dependency tree rooted by the window version of \textsf{w$_{1}$}. If a leaf vertex is a consumption group $\mathit{CG}$, two window versions of \textsf{w$_{3}$} are created and attached (a version for completion of $\mathit{CG}$, and a version for abandoning of $\mathit{CG}$); if a leaf vertex is a window version, one window version of \textsf{w$_{3}$} is created and attached.

\emph{Consumption group created.} Recall, that when a consumption group $\mathit{CG}$ associated with a window version $\mathit{WV}$ is created, the old subtree of v($\mathit{WV}$) is linked by v($\mathit{CG}$)'s abandon edge, while a modified copy of the subtree is linked by v($\mathit{CG}$)'s completion edge (Figure \ref{alg:pathtree}, lines 12--16). In the example in Figure \ref{fig:version_tree_complete}, $\mathit{WV}_2$ creates $\mathit{CG}_3$. Then, v($\mathit{CG}_3$) is attached as a new child to v($\mathit{WV}_2$), and the former child, v($\mathit{WV}_{6}$), becomes the root of the unmodified subtree of v($\mathit{CG}_3$). For all window versions in the unmodified subtree of v($\mathit{CG}_3$), a new alternative version is created that assumes that $\mathit{CG}_3$ will be completed. Suppose $\mathit{CG}_3$ contains event $E_4$. Then, window version $\mathit{WV}_{6}$ (from the unmodified subtree) contains event $E_4$, whereas the alternative window version $\mathit{WV}_{7}$ (from the modified subtree) suppresses event $E_4$.

\emph{Consumption group completed / abandoned.} When a consumption group is completed or abandoned, the respective opposite abandon or completion path of that consumption group is removed from the dependency tree. There are two different reasons why a consumption group is abandoned: (1) Due to the termination of the corresponding window version/end of window, or (2) due to a condition from a negation statement being fulfilled. For instance, a pattern specification of a sequence of events of type $A$ and $B$ can define that no event of type $C$ shall occur between the $A$ and $B$ events. If a consumption group is opened with an $A$ event, the occurrence of a $C$ event would trigger the consumption group to be abandoned as the pattern instance cannot be completed any more, even if a $B$ event would occur later. The algorithms for subtree removal are listed in Figure \ref{alg:pathtree}, lines 18--26.

\textbf{Discussion:}
To be able to process $k$ window versions in parallel we obviously need $k$ operator instances. That means, that typically only a small fraction of all possible window versions can be considered for speculative processing. To be able select the $k$ most promising window versions, we need a method for predicting the probability of possible window versions to survive (i.e., not to be dropped). In Section \ref{sec:scheduling}, we propose a scheme for scheduling the $k$ most promising  window versions on a collection of $k$ operator instances. 

\begin{figure}
\setbox0\vbox{\small
{\fontsize{8.0}{9.0}\selectfont
\begin{algorithmic}[1]
\algsetblockdefx[function]{func}{endfunc}{}{0.2cm}[3]{#1 \textbf{#2} (#3) \textbf{begin}}{\textbf{end function}}

\func{}{newWindow}{ }
	\For{\textbf{each} $\mathit{leafVertex} \in \mathit{dependency tree}$}
		\If{$\mathit{leafVertex}$ \textbf{is  window version}}
			\State $\mathit{leafVertex.child} \gets$ \textbf{new} v($\mathit{WV}$)
		\Else \Comment{else, it is a Consumption Group}
			\State $\mathit{leafVertex.completionEdge} \gets$ \textbf{new} v($\mathit{WV}$)
			\State $\mathit{leafVertex.abandonEdge} \gets$ \textbf{new} v($\mathit{WV}$)
		\EndIf
	\EndFor
\endfunc
\State
\func{}{consumptionGroupCreated}{CGroup $\mathit{CG}$, WinVersion $\mathit{WV}$}
	\State create a modified copy of the subtree attached to v($\mathit{WV}$)
	\State v($\mathit{CG}$)$.\mathit{completionEdge} \gets  $v($\mathit{WV}$)$.\mathit{modifiedSubtree}$
	\State v($\mathit{CG}$)$.\mathit{abandonEdge} \gets $v($\mathit{WV}$)$.\mathit{originalSubtree}$
\endfunc
\State

\func{}{consumptionGroupCompleted}{CGroup $\mathit{CG}$}
	\State v($\mathit{CG}$)$.\mathit{abandonEdge} \gets \mathbf{null}$
	\State v($\mathit{CG}$)$.\mathit{parent.child} \gets $v($\mathit{CG}$)$.\mathit{completionEdge}$
\endfunc
\State
\func{}{consumptionGroupAbandoned}{CGroup $\mathit{CG}$}
	\State v($\mathit{CG}$)$.\mathit{completionEdge} \gets \mathbf{null}$
	\State v($\mathit{CG}$)$.\mathit{parent.child} \gets $v($\mathit{CG}$)$.\mathit{abandonEdge}$
\endfunc

\end{algorithmic}
}
}
\centerline{\fbox{\box0}}
\vspace{-0.2cm}
\caption{Algorithms for managing the dependency tree.}
\label{alg:pathtree}
\vspace{-0.3cm}
\end{figure}

\subsection{Selecting and Scheduling the Top-k Window Versions}
\label{sec:scheduling}

The intuition behind SPECTRE is to predict the $k$ ``best'' speculative window versions and schedule them for parallel processing on $k$ operator instances. To determine the top-$k$ window versions, SPECTRE periodically determines the $k$ window versions with the highest probability to survive in the entire dependency tree. In other words, SPECTRE does not create and schedule windows, as assumed in Section \ref{sec:parallelization}, but window versions; in doing so, multiple versions of the same window can be scheduled to different operator instances in parallel.

Whether or not a window version $\mathit{WV}$ survives depends on the outcome of the preceding consumption groups, i.e. the consumption groups on the path from $\mathit{WV}$ to the root of the dependency tree. In the following, we will denote this path as $\mathit{WV}$'s \emph{root path}. Remember, each vertex representing a consumption group has two outgoing edges, a complete and an abandon edge. We say that the complete or abandon edge of a consumption group, say $\mathit{CG}$, becomes \emph{valid} when $\mathit{CG}$ is completed or abandoned, respectively. Once one of these edges becomes valid, the other one turns \emph{invalid}. Consequently, $\mathit{WV}$ survives only if all abandon and complete edges on its root path eventually become valid, i.e., $\mathit{WV}$ is dropped if at least one of these edges turn invalid.  

The probability of $\mathit{WV}$ to survive depends on the completion probabilities of the consumption groups on $\mathit{WV}$'s root path. The survival probability of $\mathit{WV}$, denoted as $\mathit{SP(WV)}$ is determined as follows: Let $\mathit{P(CG)}$ be the probability that $\mathit{CG}$ is completed. Moreover, let $\mathit{CG}_c$ and $\mathit{CG}_a$ be the set of consumption groups that contribute a complete and abandon edge to $\mathit{WV}$'s root path, respectively. Then\footnote{Note that this calculation bases on the assumption that the different consumption groups are completed or abandoned independently from each other. If there are dependencies between different occurrences of a pattern and, hence, between the completion of different consumption groups, this can be incorporated in the probability calculation by using dependent / conditional probabilities. However, for the sake of simplicity of the presentation of technical concepts and algorithms, we use the formula for independent probabilities here.}, $\mathit{SP(WV)} = \prod_{c \in \mathit{CG}_c} P(c) \times \prod_{c' \in \mathit{CG}_a}(1 - P(c'))$.

\subsubsection{Prediction Model} 

Now, we discuss how we predict the completion probability  of a consumption group. 
Generally, we observe that the probability that a consumption group is completed equals to the probability that the underlying partial match for a search pattern is completed.
Our scheme for predicting the completion probability $\mathit{P(CG)}$ of a consumption group $\mathit{CG}$ at a given time takes into account two factors:  (1) The inverse degree of completion, i.e., how many more events are at least required in order to complete the pattern---denoted by $\delta$---and (2) the expected number of events left in the window, denoted by $n$. If $\delta$ is low and many events are still expected to occur in the window, the probability of completion is high. On the other hand, if $\delta$ is high and only very few events are still expected in the window, the probability of completion is low. In the following, we describe how the probabilistic model is built and updated at system run-time.

The dynamic process of pattern completion while processing events is modeled as a discrete-time \emph{Markov process}.  The state of the Markov process is spanned from $\delta$ to 0. For instance, if a pattern instance consists of at least 3 events (e.g., a sequence of 3 events, or a set of 3 events), the state-space has the elements ``3'', ``2'', ``1'' and ``0'', with ``0'' representing the state of total pattern completion.
Based on statistics monitored at system run-time, a \emph{stochastic matrix} $T_1$ is built that describes the \emph{transition probabilities} between the states of the Markov process when processing \emph{one} event. To this end, window versions of independent windows gather statistics about the probability of changing from $\delta_{\mathit{old}}$ to $\delta_{\mathit{new}}$ when an event is processed. The transition probabilities between any pair of $\delta_{\mathit{old}}$ and $\delta_{\mathit{new}}$ are captured in a matrix $T_1^{new}$. After $\rho$ new measurements are available, an updated $T_1$ is computed from the old $T_1^{old}$ and the newly calculated $T_1^{new}$ as  $T_1 = (1 - \alpha) * T_1^{old} + \alpha * T_1^{new}$ (\emph{exponential smoothing}). $\alpha \in [0,1]$ is a system parameter to control the impact of recent and of old statistics on $T_1$.

\begin{figure}
\setbox0\vbox{\small
{\fontsize{8.0}{9.0}\selectfont
\begin{algorithmic}[1]
\algsetblockdefx[function]{func}{endfunc}{}{0.2cm}[3]{#1 \textbf{#2} (#3) \textbf{begin}}{\textbf{end function}}

\func{}{predictCompletionProbability}{ConsumptionGroup $\mathit{CG}$}
	\State $n \gets \mathit{Splitter.avgWindowSize - posInWindow}$
	\If{$n \leq 0$}
	\State $n \gets 1$ \Comment{At least 1 more event expected}
	\EndIf
	\State $T_n \gets (1 - \frac{n \bmod \ell}{\ell}) * T_{\lfloor \frac{n}{\ell} \rfloor * \ell} + \frac{n \bmod \ell}{\ell} * T_{\lceil \frac{n}{\ell} \rceil * \ell}$
	\State $\delta \gets \mathit{CG.completionState}$
	\State $v_0 \gets \delta$-th unit vector
	\State $v_n \gets T_n * v_0$
	\State $\mathbf{return}\ v_n[\mathit{last}]$
\endfunc

\end{algorithmic}
}
}
\centerline{\fbox{\box0}}
\vspace{-0.2cm}
\caption{Calculation of completion probability of a consumption group. }
\label{alg:cgroup}
\vspace{-0.3cm}
\end{figure}

Now, the probability of state transitions when processing $n$ events can be computed by raising $T_1$ to the $n$-th power:  $T_n = (T_1)^n$. The initial state is modeled as a row vector $v_0 = (0, ..., 0, 1, 0, ..., 0)$---the $\delta$-th unit vector, where the $\delta$-th position is 1 and all other positions are 0. The probabilities of reaching the different states in $n$ steps can be computed as $v_n = T_n * v_0$. The last entry of $v_n$, referring to state ``0'', is the probability to complete the pattern in $n$ steps starting from state $v_0$.

To reduce the number of matrix multiplications, each time when $T_1$ is updated, a set of predefined ``step sizes'' is precomputed, e.g.,  $T_{10}$, $T_{20}$, $T_{30}$, etc., providing transition probabilities when 10, 20, 30, ... events are processed. If the number of expected events $n$ is in between two precomputed steps, the transition probabilities are linearly interpolated, e.g., $T_{14} = 0.6 * T_{10} + 0.4 * T_{20} $. The step size, denoted as $\ell$, is a system parameter. 

Figure \ref{alg:cgroup} formalizes the described methods in an algorithm. The expected number of events left in the window, $n$, is calculated from the average window size monitored in the splitter and the position of the last processed event in the window (line 2). The probability matrix $T_n$ is calculated by linear interpolation of precomputed matrices (line 6). $\delta$ is obtained directly from $\mathit{CG}$ (line 7), and used in order to build $v_0$ (line 8); $v_n$ is calculated according to the description above (line 9). The resulting completion probability (transition to state `` 0'' / pattern completed) is returned (line 10).

\subsubsection{Scheduling}
Here, we describe how SPECTRE periodically selects and schedules the k window versions with the highest survival probability. 

Notice that the survival probability of window versions is decreasing in a root-to-leaf direction in the dependency tree, i.e. in a window version's subtree there exist only window versions that have the same or a lower survival probability. Therefore, window versions are already sorted by their survival probability in the dependency tree, so that it already represents a max-heap, which simplifies the selection of the top-$k$ versions substantially. From top to the bottom, window versions are added to the top-$k$ list as detailed in the algorithm in Figure \ref{alg:topk}. The algorithm works with two data structures: (1) a set storing the resulting top-$k$ versions (line 2), and (2) a priority queue storing candidates for being added to the top-$k$ versions (line 3). The priority queue sorts the contained versions by their probability, highest probability first. Until $k$ versions are found, the highest version from the candidate list is added to the result set (lines 4--6). The children of that version are also added as candidates (lines 7--9). This way, the top-$k$ window versions are determined with only visiting the minimal number of vertices in the dependency tree. 

The scheduling algorithm, listed in Figure \ref{alg:scheduling}, does not re-schedule window versions that are already scheduled to avoid unnecessary operations and to increase memory and cache locality of operator instances. Hence, the to-be-scheduled versions are determined (lines 7--9). Further, ``free'' operator instances are determined that will get a new window version scheduled (lines 10--11). Then, every window version that needs to be scheduled is scheduled to one of the free operator instances (lines 14--17).

\begin{figure}
\setbox0\vbox{\small
{\fontsize{8.0}{9.0}\selectfont
\begin{algorithmic}[1]
\algsetblockdefx[function]{func}{endfunc}{}{0.2cm}[3]{#1 \textbf{#2} (#3) \textbf{begin}}{\textbf{end function}}

\func{}{findTopKVersions}{$\mathit{dependencyTree}$, $k$}
	\State $\mathit{result} \gets \{\}$ \Comment{set}
	\State $\mathit{candidates} \gets \{\mathit{dependencyTree.root}\}$ \Comment{priority queue}
	\For{$i\gets1 ... k$}
		\State $\mathit{tmp} \gets \mathit{candidates.pop()}$
		\State $\mathit{result.append(tmp)}$
		\For{\textbf{each} $M\gets \mathit{tmp.child}$} 
			\State $\mathit{candidates.add(M)}$
		\EndFor
	\EndFor
	\State \textbf{return} $\mathit{result}$
\endfunc

\end{algorithmic}
}
}
\centerline{\fbox{\box0}}
\vspace{-0.2cm}
\caption{Top-k window version selection algorithm.}
\label{alg:topk}
\vspace{-0.2cm}
\end{figure}

\begin{figure}
\setbox0\vbox{\small
{\fontsize{8.0}{9.0}\selectfont
\begin{algorithmic}[1]
\algsetblockdefx[function]{func}{endfunc}{}{0.2cm}[3]{#1 \textbf{#2} (#3) \textbf{begin}}{\textbf{end function}}

\State List$\langle$OperatorInstance$\rangle$ $\mathit{operatorInstances}$
\State Tree $\mathit{dependencyTree}$

\func{}{schedule}{ }
	\State List$\langle$WindowVersion$\rangle$ $\mathit{toBeScheduled}$ \Comment{empty list} 
	\State List$\langle$OperatorInstance$\rangle$ $\mathit{freeOperatorInstances} \gets \mathit{operatorInstances}$  
	\State List$\langle$WindowVersion$\rangle$ $\mathit{topkVersions}$ $\gets \mathtt{findTopKVersions(\mathit{dependencyTree})}$
	\For{\textbf{each} WindowVersion $\mathit{WV}$ in $\mathit{topkVersions}$} \Comment{first pass}
		\If{not $\mathit{WV}$.isScheduled()} \Comment{WV must be scheduled}
			\State $\mathit{toBeScheduled}.\mathtt{add(\mathit{WV})}$
		\Else \Comment{the operator instance keeps WV}
			\State $\mathit{freeOperatorInstances}.\mathtt{remove(\mathit{WV}.getOperatorInstance())}$
		\EndIf
	\EndFor
	\For{\textbf{each} WindowVersion $\mathit{WV}$ in $\mathit{toBeScheduled}$} \Comment{second pass}
		\State OperatorInstance $\mathit{OP} \gets \mathit{freeOperatorInstances}.\mathtt{pop()}$ 
		\State $\mathit{OP.scheduledWV} \gets \mathit{WV}$
	\EndFor
\endfunc

\end{algorithmic}
}
}
\centerline{\fbox{\box0}}
\vspace{-0.2cm}
\caption{Splitter: Scheduling algorithm.}
\label{alg:scheduling}
\vspace{-0.3cm}
\end{figure}

\subsection{Parallel Processing of Window Versions}
\label{sec:processing}
Here, we describe how operator instances process their assigned window version according to the dependencies in the dependency tree. In particular, we describe how events are processed and suppressed, and how consumption groups are updated when sub-patterns are detected in a window version.


\begin{figure}
\setbox0\vbox{\small
{\fontsize{8.0}{9.0}\selectfont
\begin{algorithmic}[1]
\algsetblockdefx[function]{func}{endfunc}{}{0.2cm}[3]{#1 \textbf{#2} (#3) \textbf{begin}}{\textbf{end function}}
\State WindowVersion $\mathit{currentWV}$ \Comment{currently processed WV}
\State WindowVersion $\mathit{scheduledWV}$ \Comment{currently scheduled WV}
\State int $i \gets 0$ \Comment{processing counter}

\func{}{main}{}
	\While{true}
		\State $i \gets i + 1$
		\If{$\mathit{scheduledWV} \neq \mathit{currentWV}$} \Comment{changed WV?}
			\State $\mathit{currentWV} \gets \mathit{scheduledWV}$
		\EndIf
		
		\State
		\State \emph{// process the next event}
		\State Event $\mathit{nextEvent} \gets \mathit{currentWV.Window.getNextEvent()}$ 
		\If{$\mathit{nextEvent}$ not in $\mathit{currentWV.suppressedCGs}$}
			\State Feedback $\mathit{fb} \gets$ process($\mathit{nextEvent}$)
			\If{$\mathit{fb}$: emitted complex event $E$, completed $\mathit{CG_c}$}
				\State buffer $E$
				\State dependencyTree.$\mathit{consumptionGroupCompleted(CG_c)}$
			\EndIf
			\If{$\mathit{fb}$: abandoned $\mathit{CG_{a}}$}
				\State dependencyTree.$\mathit{consumptionGroupAbandoned(CG_{a})}$
			\EndIf
			\If{$\mathit{fb}$: created $\mathit{CG_{new}}$}
				\State dependencyTree.
				\State \ \ \ \ $\mathit{consumptionGroupCreated(CG_{new}, currentWV)}$
			\EndIf
			\If{$\mathit{fb}$: added $\mathit{nextEvent}$ to $\mathit{CG}$}
				\State $\mathit{CG.add(nextEvent)}$
			\EndIf
		\EndIf
	
		\State
		\State \emph{// consistency check after each $i$ steps}
		\If{$(i$ mod $\mathit{consistencyCheckFreq}) == 0$} \Comment{consistency check}
			\State bool $\mathit{inconsistencyDetected} \gets$ false
			\For{$\mathit{CG} \in \mathit{currentWV.suppressedCGs}$}    
				\If{$\mathit{CG.version} != \mathit{CG.lastCheckedVersion}$} 
					\If{$\mathit{currentWV.usedEvents} \cap \mathit{CG.events} \neq \emptyset$}
						\State $\mathit{inconsistencyDetected} \gets$ true
					\EndIf
				\EndIf
				\State $\mathit{CG.lastCheckedVersion} \gets \mathit{CG.version}$
			\EndFor
			\If{$\mathit{inconsistencyDetected}$}
				\State rollback $\mathit{currentWV}$
			\EndIf
		\EndIf \Comment{end of consistency check}

	\EndWhile
	
\endfunc

\end{algorithmic}
}
}
\centerline{\fbox{\box0}}
\vspace{-0.2cm}
\caption{Operator Instances: Event Processing. }
\label{alg:instance}
\vspace{-0.3cm}
\end{figure}

The scheduled window versions are processed in parallel by the associated operator instances. This means, that an operator instance processes or suppresses events according to the dependencies of the window version. In particular, when the root path of the window version meets the completion edge of a consumption group, events in that consumption group are not processed: they are suppressed.  Complex events produced when processing a speculative window version are kept buffered until the window version either becomes valid---then, the complex events are emitted---or is dropped---then, the complex events are dropped, too.  Further, when an event is processed, updates of the consumption groups can occur (creation, completion or abandoning a consumption group, or adding the event to an existing consumption group). In the following, we detail the underlying algorithms.

Figure \ref{alg:instance} lists the algorithm for event processing in the operator instances. In the beginning of a processing cycle, the operator instance checks whether the splitter has scheduled a new window version (lines 7--9). Then, the next event of the currently scheduled window version is processed (lines 11--29). The operator instance checks whether the event is part of any consumption group that shall be suppressed (line 13). If this is the case, the event is suppressed, i.e., its processing is skipped. If the event is not suppressed, it is processed according to the operator logic (line 14). In doing so, there can be four different actions triggered based on feedback the operator logic provides. (1) The processed event can complete one or multiple partial matches: This induces the creation of one or multiple complex events and the completion of the associated consumption groups. In that case, the emitted complex events are buffered, and the dependency tree is updated, calling the \emph{consumptionGroupCompleted} function (cf. Section \ref{sec:speculation}). (2) The processed event can lead to the abandoning of consumption groups, either by closing the window, or by invalidating the underlying partial match. In this case, the dependency tree is updated, calling the \emph{consumptionGroupAbandoned} function (cf. Section \ref{sec:speculation}). (3) The processed event can lead to the creation of a new consumption group by initiating a new partial match. In this case, the dependency tree is updated, calling the \emph{consumptionGroupCreated} function (cf. Section \ref{sec:speculation}). (4) The processed event can become part of one or several existing partial matches, possibly adding the event to the associated consumption groups. In this case, the affected consumption groups are updated directly without changing the structure of the dependency tree. Note, that in the implementation of SPECTRE, the function calls of the operator instances on the dependency tree are buffered---they are actually executed on the dependency tree in a batch at each new scheduling cycle of the splitter.

The $k$ scheduled window versions are processed concurrently by the $k$ operator instances, without synchronizing the processing progress of the different window versions. This can lead to a situation where an update on an existing consumption group is propagated too late, causing inconsistencies. For instance, when an event is added to a consumption group  $\mathit{CG}$ in one window version $\mathit{WV_a}$ after it has been processed in another window version $\mathit{WV_b}$ adjacent to $\mathit{CG}$'s completion edge, an inconsistency can be induced in $\mathit{WV_b}$ (i.e., an event is processed that should be suppressed). To detect such situations, SPECTRE employs periodic \emph{consistency checks}; the underlying algorithm is sketched in lines 31 -- 45. For every consumption group to be suppressed in the currently processed window version, the algorithm checks whether an update has occurred since the last consistency check. If this is the case, the algorithm checks whether in the current window version, any event in the updated consumption group has been erroneously processed. If yes, then an inconsistency has been detected: The event should have been suppressed, but has actually been processed. If an inconsistency is detected, the state of the window version is rolled back to the start, i.e., the window version is reprocessed from the start. Instead of reprocessing a window version from the start in case of an inconsistency, it could also be recovered from an intermediate checkpoint. However, when implementing that approach, we realized that the overhead in periodically checkpointing all window versions is much higher than the gain from recovering from checkpoints.

\section{Evaluations}
\label{sec:Evaluations}

In this section, we evaluate the performance of SPECTRE under different real-world and synthetic workloads and varying queries in the setting of an algorithmic trading scenario. We analyze the scalability of SPECTRE with a growing number of operator instances and the overhead involved in speculation and dependency management. 

\subsection{Experimental Setup}
Here, we describe the evaluation platform, the SPECTRE implementation and the datasets and queries used in the evaluations.

\textbf{Evaluation Platform.}
We run SPECTRE on a shared memory multi-core machine with 2x10 CPU cores (Intel Xeon E5-2687WV3 3.1 GHz) that support hyper-threading (i.e., 40 hardware threads). The total available memory in the machine is 128 GB and the operating system is CentOS 7.3.

\textbf{Implementation.}
SPECTRE is implemented using C++.  The pattern detection and window splitting logic of the queries in these evaluations are implemented as a user-defined function (UDF) inside SPECTRE. Further, we provide a client program that reads events from a source file and sends them to SPECTRE over a TCP connection. Our implementation of SPECTRE is open source\footnote{https://github.com/spectreCEP}.

\textbf{Datasets.}
We employ two different datasets centered around an algorithmic trading scenario. 

First, a real-world stock quotes stream originating from the New York Stock Exchange (NYSE). 
This dataset contains real intra-day quotes of around 3000 stock symbols from NYSE collected over two months from Google Finance\footnote{https://www.google.com/finance}; in total, it contains more than 24 million stock quotes. The quotes have a resolution of 1 quote per minute for each stock symbol. We refer to this dataset as the \emph{NYSE Stock Quotes} dataset, denoted as \emph{NYSE}. NYSE represents realistic data for stock market pattern analytics.

Second, we generated a random sequence of 3 million events consisting of 300 different stock symbols; the probability of each stock symbol is equally distributed in the sequence. We refer to this dataset as the \emph{Random Stock Symbols} dataset, denoted as \emph{RAND}.

\begin{figure}
\begin{minipage}[t]{0.42\linewidth}
	\begin{equation*}[Q1]
\scalebox{0.59}{$
\begin{aligned} 
\ & \mathtt{\textbf{PATTERN}\ (MLE ~RE_1 ~RE_2 ~... ~RE_q)} \\
\ & \mathtt{\quad \textbf{DEFINE}} \\
\ & \mathtt{\qquad MLE ~\textbf{AS}~(MLE.closePrice}\\ 
\ & \mathtt{\qquad\qquad > MLE.openPrice),}\\ 
\ & \mathtt{\qquad RE_1 ~\textbf{AS}~(RE_1.closePrice } \\
\ & \mathtt{\qquad\qquad  > RE_1.openPrice),}\\ 
\ & \mathtt{\qquad RE_2 ~\textbf{AS}~(RE_2.closePrice} \\
\ & \mathtt{\qquad\qquad   > RE_2.openPrice),}\\ 
\ & \mathtt{\qquad ..., } \\
\ & \mathtt{\qquad RE_q ~\textbf{AS}~(RE_q.closePrice} \\
\ & \mathtt{\qquad\qquad    > RE_q.openPrice)  }\\ 
\ & \mathtt{\quad\textbf{WITHIN}\ \textit{ws}\ events\ \textbf{FROM}\ MLE} \\
\ & \mathtt{\quad\textbf{CONSUME}\ (MLE~RE_1~RE_2~...~RE_q)} \\
\end{aligned}
$}
\end{equation*}

\vspace{1cm}

\begin{equation*}[Q3]
\scalebox{0.59}{$
\begin{aligned} 
\ & \mathtt{\textbf{PATTERN}\ (A ~SET(~X_1 ~... ~X_n))} \\
\ & \mathtt{\quad\textbf{WITHIN}\ \textit{ws}\ events\ }\\
\ & \mathtt{\qquad\qquad \textbf{FROM}\ every\ s\ events } \\
\ & \mathtt{\quad\textbf{CONSUME}\ (A ~SET(~X_1 ~... ~X_n))} \\
\end{aligned}
$}
\end{equation*}

\end{minipage}%
    \hfill%
\begin{minipage}[t]{0.56\linewidth}
	\begin{equation*}[Q2]
\scalebox{0.59}{$
\begin{aligned} 
\ & \mathtt{ \textbf{PATTERN}\ (A ~B^+ ~C ~D^+ ~E ~F^+ ~G ~H^+ ~I ~J^+ ~K ~L^+ ~M)} \\
\ & \mathtt{\quad \textbf{DEFINE}} \\
\ & \mathtt{\qquad A~\textbf{AS}~(A.closePrice < lowerLimit),  } \\
\ & \mathtt{\qquad B~\textbf{AS}~(B.closePrice > lowerLimit} \\
\ & \mathtt{\qquad \qquad \textbf{AND}~ B.closePrice < upperLimit),} \\
\ & \mathtt{\qquad C~\textbf{AS}~(C.closePrice > upperLimit),  } \\
\ & \mathtt{\qquad D~\textbf{AS}~(D.closePrice > lowerLimit} \\
\ & \mathtt{\qquad \qquad \textbf{AND}~ D.closePrice < upperLimit),} \\
\ & \mathtt{\qquad E~\textbf{AS}~(E.closePrice < lowerLimit),  } \\
\ & \mathtt{\qquad F~\textbf{AS}~(F.closePrice > lowerLimit} \\
\ & \mathtt{\qquad \qquad \textbf{AND}~ F.closePrice < upperLimit),} \\
\ & \mathtt{\qquad G~\textbf{AS}~(G.closePrice > upperLimit),  } \\
\ & \mathtt{\qquad H~\textbf{AS}~(H.closePrice > lowerLimit} \\
\ & \mathtt{\qquad \qquad \textbf{AND}~ H.closePrice < upperLimit),} \\
\ & \mathtt{\qquad I~\textbf{AS}~(G.closePrice < lowerLimit),  } \\
\ & \mathtt{\qquad J~\textbf{AS}~(J.closePrice > lowerLimit} \\
\ & \mathtt{\qquad \qquad \textbf{AND}~ J.closePrice < upperLimit),} \\
\ & \mathtt{\qquad K~\textbf{AS}~(G.closePrice > upperLimit),  } \\
\ & \mathtt{\qquad L~\textbf{AS}~(L.closePrice > lowerLimit} \\
\ & \mathtt{\qquad \qquad \textbf{AND}~ L.closePrice < upperLimit),} \\
\ & \mathtt{\qquad M~\textbf{AS}~(M.closePrice < lowerLimit),  } \\
\ & \mathtt{\quad\textbf{WITHIN}\ \textit{ws}\ events\ \textbf{FROM}\ every\ s\ events} \\
\ & \mathtt{\quad\textbf{CONSUME}\ (A ~B^+ ~C ~D^+ ~E ~F^+ ~G ~H^+ ~I ~J^+ ~K ~L^+ ~M)} \\
\end{aligned}
$}
\end{equation*}
\end{minipage}%
    \hfill%
	
\vspace{-0.2cm}
\caption{Queries.}
\label{fig:queries}
\vspace{-0.3cm}
\end{figure}

\textbf{Queries.}
We employ three different queries, Q1 to Q3, in the evaluations (cf. Figure \ref{fig:queries}). The queries are listed in the \texttt{MATCH-RECOGNIZE} notation \cite{zemke2007pattern}, which is concise and easy to understand. Note, that we extended the \texttt{MATCH-RECOGNIZE} notation by two additional constructs stemming from the Tesla language \cite{Cugola:2010:TFD:1827418.1827427}: \texttt{WITHIN ... FROM} to specify a window size and window start condition, and \texttt{CONSUME} to specify consumption policies.

Q1 detects a complex event when the first $q$ rising or the first $q$ falling stock quotes of any stock symbol (defined as $\mathtt{RE}$ or $\mathtt{FE}$, respectively) are detected within $\mathit{ws}$ minutes from a rising or falling quote of a leading stock symbol (defined as $\mathtt{MLE}$). The leading stock symbols are composed of a list of 16 technology blue chip companies. In the listing of Q1, we show only the stock rising pattern; the falling pattern is constructed accordingly. In case a complex event is detected, all constituent incoming events are consumed. Note, that this query always has a fixed pattern length of $q$, and each matching event moves the pattern detection to a higher completion stage.

Q2 is a query from related work (Balkesen and Tatbul \cite{Balkesen:2013:RRI:2488222.2488257}, Query 9) that we extended by a window size of $\mathit{ws}$ events, a window slide of $s$ events and a consumption policy. It detects a complex event when specific changes occur in the price of a stock symbol between defined \textit{upper} and \textit{lower} limits.  As in Q1, all constituent incoming events are consumed when a complex event is detected. We use the \textit{lower} and \textit{upper} limits to control the average pattern size. A small \textit{lower} and a large \textit{upper} limit results in a larger average pattern size, and vice versa. 
In contrast to Q1, Q2 has a variable length even for a fixed \textit{lower} and \textit{upper} limit. A matching event might or might not influence the pattern completion: the Kleene$^+$ implies that many events can match while the pattern completion does not progress.

Q3 detects a set of $n$ specific stock symbols following stock symbol $A$. In contrast to the other queries, the ordering of those $n$ symbols is not important. The pattern length $n$, window size $\mathit{ws}$, and window slide $s$ can be freely varied. All constituent events are consumed when a complex event is detected.

\subsection{Performance Evaluation}

In this section, we evaluate the throughput and scalability of SPECTRE. 
First of all, we evaluate how SPECTRE performs with a growing number of parallel operator instances and with different consumption group completion probabilities. After that, we provide a detailed analysis of the Markov model SPECTRE uses to predict the completion probability of consumption groups. Finally, we discuss a comparison to the CEP engine T-REX \cite{Cugola:2012:CEP:2221990.2222305}.

If not noted otherwise, we employ the following settings. The number of created consumption groups is limited to one per window version. The Markov model is employed with the parameters $\alpha = 0.7$ and $\ell = 10$.

 To measure the system throughput, we streamed the datasets as fast as possible to the system. Each experiment was repeated 10 times. The figures show the 0th, 25th, 50th, 75th and 100th percentiles of the experiment results in a ``candlesticks'' representation.

\subsubsection{Scalability}

\begin{figure*}
\begin{minipage}[t]{0.3\linewidth}
	\begin{overpic}[width=\linewidth]{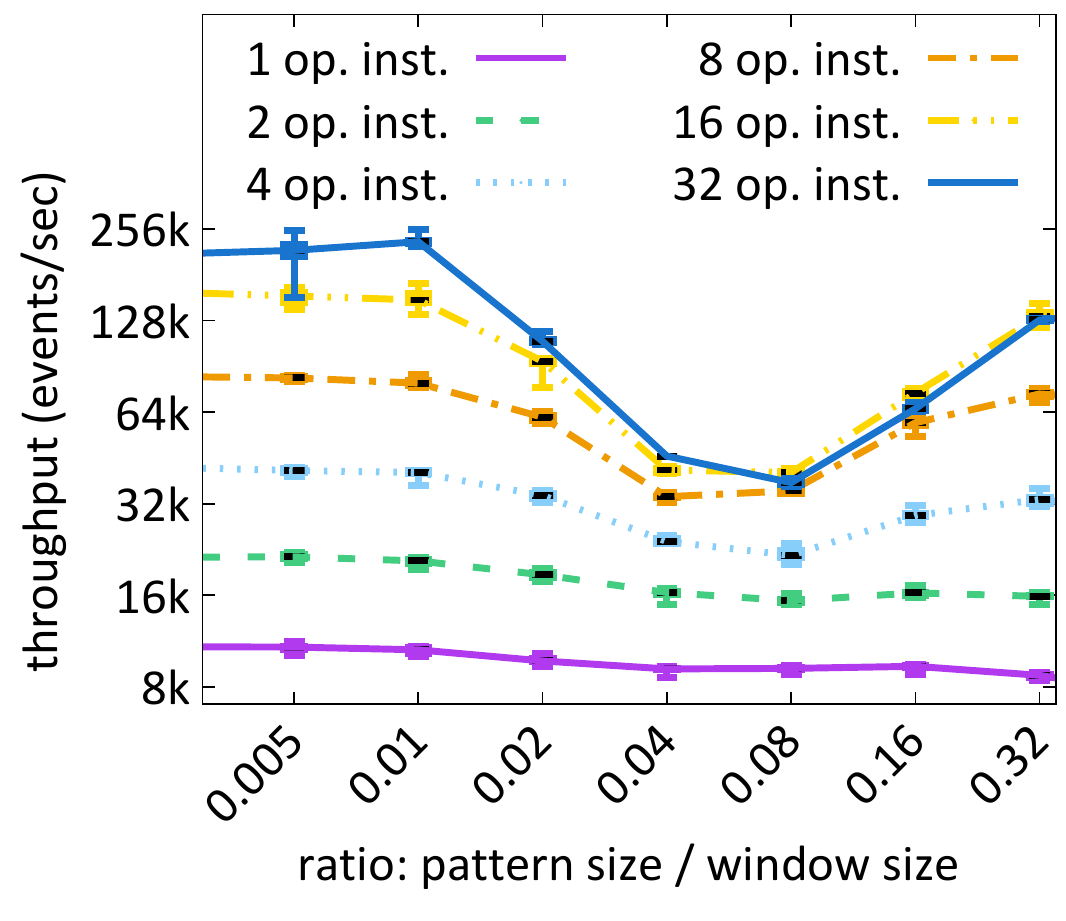}
        		\put(0,0){(a)}
      	\end{overpic}
\end{minipage}%
    \hfill%
\begin{minipage}[t]{0.3\linewidth}
	\begin{overpic}[width=\linewidth]{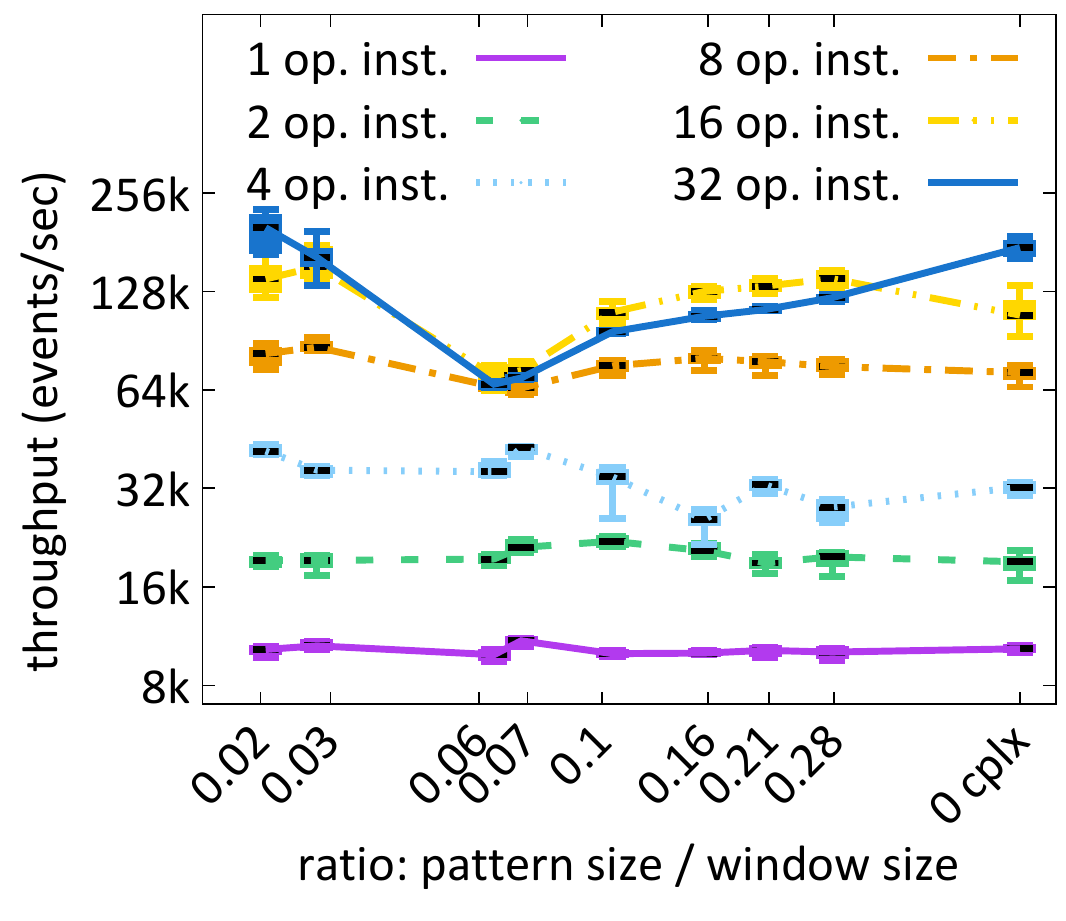}
        		\put(0,0){(b)}
      	\end{overpic}
\end{minipage}%
    \hfill%
\begin{minipage}[t]{0.3\linewidth}
	\begin{overpic}[width=\linewidth]{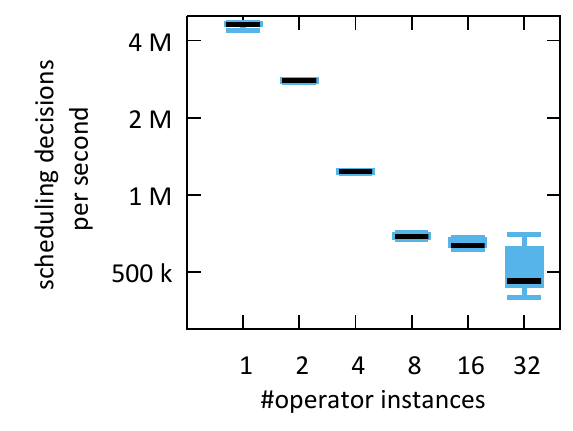}
        		\put(0,0){(c)}
      	\end{overpic}
\end{minipage}%
    \newline \newline
	\begin{minipage}[t]{0.3\linewidth}
	\begin{overpic}[width=\linewidth]{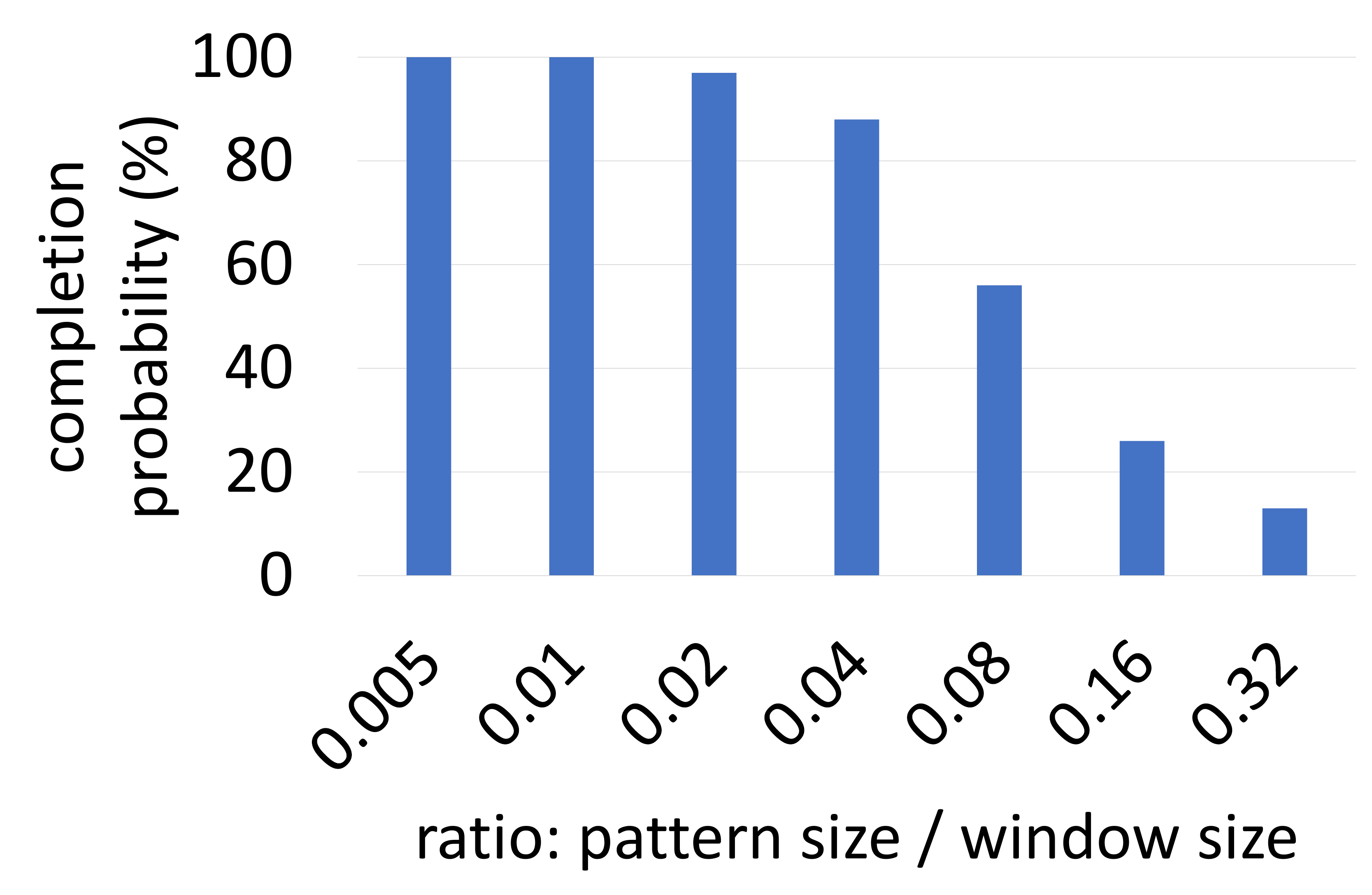}
        		\put(0,0){(d)}
      	\end{overpic}
\end{minipage}%
    \hfill%
\begin{minipage}[t]{0.3\linewidth}
	\begin{overpic}[width=\linewidth]{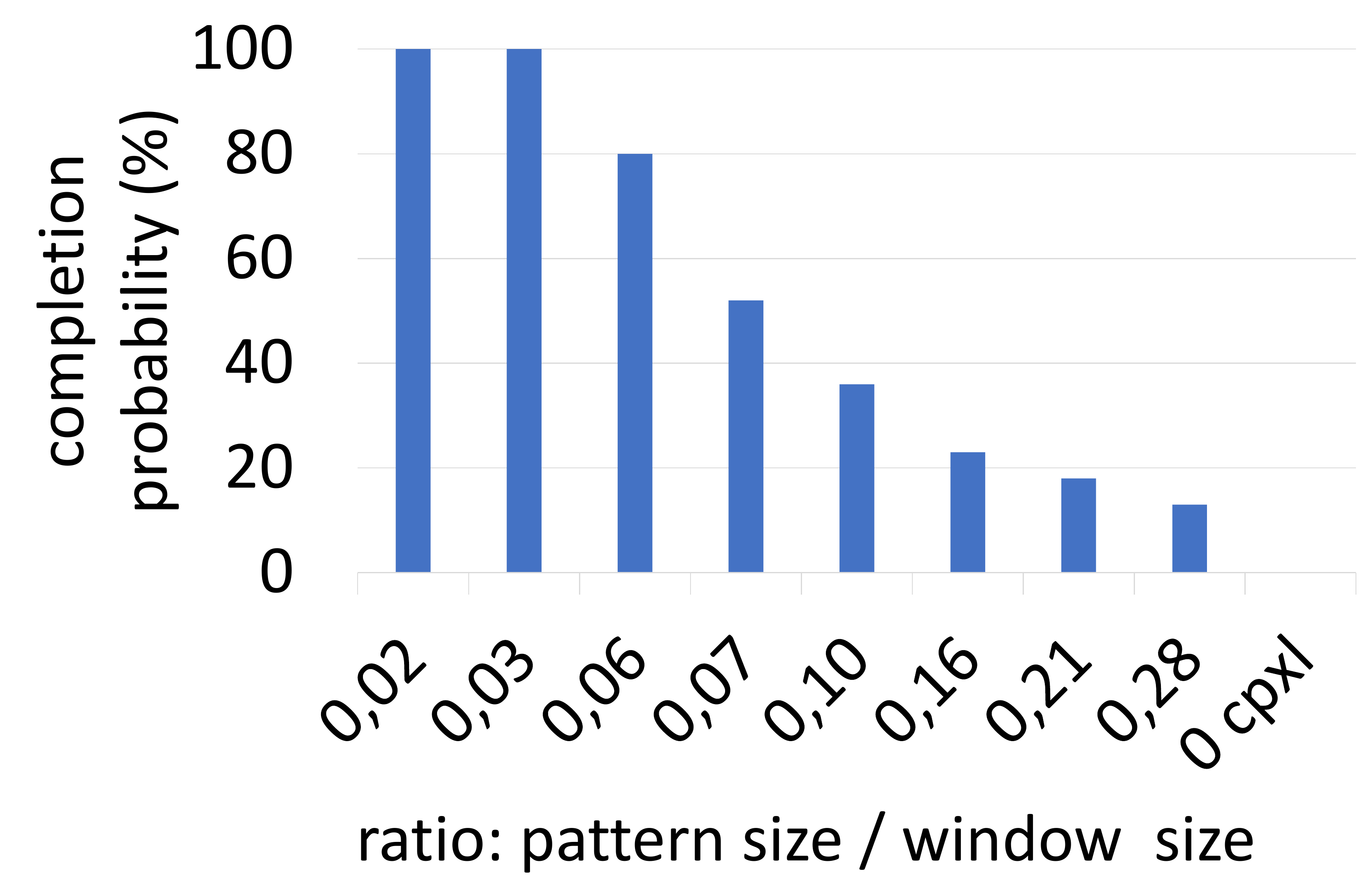}
        		\put(0,0){(e)}
      	\end{overpic}
\end{minipage}%
    \hfill%
\begin{minipage}[t]{0.3\linewidth}
	\begin{overpic}[width=\linewidth]{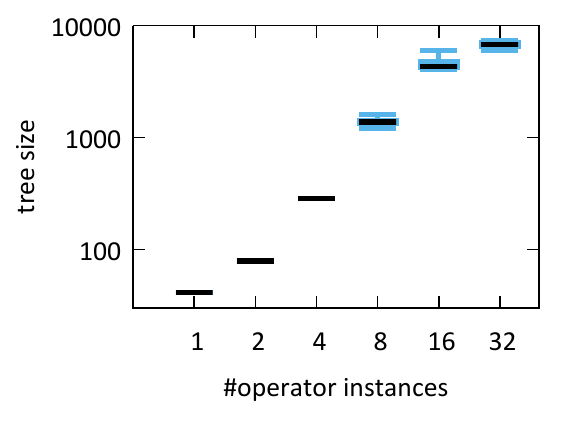}
        		\put(0,0){(f)}
      	\end{overpic}
\end{minipage}%
    \hfill%
	
\vspace{-0.1cm}
\caption{Evaluations. (a)+(d): Scalability (Q1 on NYSE). (b)+(e): Scalability (Q2 on NYSE). (c)+(f): Overhead (Q1 on NYSE).}
\label{fig:performance_worker_threads}
\vspace{-0.2cm}
\end{figure*}

Here, we evaluate the scalability of SPECTRE. To this end, we analyze the system throughput, i.e., the number of events processed per second, with a growing number of operator instances. The following questions are addressed: (1) How does the scalability depend on the completion probability of the consumption groups? (2) How much computational and memory overhead is induced by maintaining the dependency tree and determining the top-$k$ window versions?

\paragraph{Effect of Completion Probability of Consumption Groups}

We expect that the completion probability of consumption groups influences the system throughput. To make that clear, regard two extreme cases: All consumption groups are abandoned, or all consumption groups are completed. In the first case, SPECTRE should only schedule window versions on the \emph{left-most} path of the dependency tree. In the second case, SPECTRE should only schedule window versions on the \emph{right-most} path of the dependency tree. In both cases, the scheduling algorithm should traverse the dependency tree in \emph{depth}; i.e., it should schedule $k$ window versions from $k$ different windows. Further, none of the scheduled window versions should be dropped; all of them should survive. Hence, the throughput should be maximal. On the other hand, suppose that the completion probability of all consumption groups is constantly at 50 \%. In that case, SPECTRE should traverse the dependency tree in \emph{breadth}; i.e., it should schedule 1 window version of the first window, 2 window versions of the second window, 4 window versions of the third window, etc. However, only 1 window version of each window can survive; all others will be dropped. Hence, the higher $k$ is, the more futile processing is performed, as the probability to predict the correct window version drops exponentially with $k$. In the following, we analyze whether SPECTRE shows the expected behavior and discuss implications.

To this end, we run a set of experiments with queries Q1 and Q2, using the NYSE dataset. In both queries, there are parameters that can be changed such that the average completion probability of consumption groups is manipulated. In Q1, we achieve this by directly setting the pattern size $q$, such that the ratio between pattern size and window size changes. Larger patterns are less likely to complete. In Q2, we cannot directly set the pattern size. However, we influence the average pattern size---and thus, the average completion probability---by changing the upper and lower limit parameters in the pattern definition. 

In Q1, we employ a sliding window with a window size $\mathit{ws}$ of 8,000 events, setting pattern sizes $q$ of 40, 80, 160, 320, 640, 1280, and 2560 events. We calculate a ``ground truth'' value of the completion probability of consumption groups by performing a sequential pass without speculations: The number of created consumption groups divided by the number of produced complex events provides the ground truth value. The system throughput employing 1, 2, 4, 8, 16, and 32 operator instances, is depicted in Figure \ref{fig:performance_worker_threads} (a). The corresponding ground truth probabilities are depicted in Figure~\ref{fig:performance_worker_threads}~(d).

At a ratio of pattern size to window size of 40 / 8,000 (i.e., 0.005), the ground truth of consumption group completion probability is at 100 \%, i.e., all partial matches are completed. The throughput scales almost linearly with a growing number of operator instances, from 10,800 events/second at 1 operator instance to 154,000 events/second at 16 operator instances (scaling factor 14.3) and 218,000 events/second at 32 operator instances (scaling factor 20.2). 
Increasing the pattern size decreases the completion probability of consumption groups. At a ratio of pattern size to window size of 640 / 8,000 (i.e., 0.08), the ground truth of consumption group completion probability is at 56 \%, i.e., half of partial matches are completed and the other half are abandoned. The throughput scales from 9,200 events/second at 1 operator instance to 35,000 events/second at 8 operator instances (scaling factor 3.8). However, employing more than 8 operator instances does not increase the throughput further: With 16 and 32 operator instances, it is comparable to 8 operator instances. 
Further increasing the pattern size, we reach a ground truth of consumption group completion probability of 13 \% at a ratio of pattern size to window size of 2560 / 8,000 (i.e., 0.32). Here, the throughput scales better, from 8,700 events/second at 1 operator instance to 131,900 events/second at 16 operator instances (scaling factor 15.2). Here, 32 operator instances do not improve the throughput further compared to 16 operator instances. 

In Q2, we employ a sliding window with a window size $\mathit{ws}$ of 8,000 events and a sliding factor $s$ of 1,000 events. We arranged the lower and upper limit parameters in the pattern definition such that the corresponding average pattern sizes were 180, 226, 496, 560, 839, 1261, 1653, and 2223 events, plus one setting that made it impossible for a pattern to be completed.  The system throughput employing 1, 2, 4, 8, 16, and 32 operator instances, is depicted in Figure \ref{fig:performance_worker_threads} (b). The corresponding ground truth probabilities are depicted in Figure~\ref{fig:performance_worker_threads}~(e).

At a ratio of pattern size to window size of 180 / 8,000 (i.e., 0.02), the ground truth of consumption group completion probability is at 100 \%, i.e., all partial matches are completed. The throughput scales almost linearly with a growing number of operator instances, from 10,300 events/second at 1 operator instance to 139,800 events/second at 16 operator instances (scaling factor 13.8) and 200,400 events/second at 32 operator instances (scaling factor 19.5). At a ratio of pattern size to window size of 560 / 8,000 (i.e., 0.07), the ground truth of consumption group completion probability is at 50 \%, i.e., half of partial matches are completed and the other half are abandoned. The throughput scales from 10,900 events/second at 1 operator instance to 64,900 events/second at 8 operator instances (scaling factor 6.0). Employing more than 8 operator instances does not increase the throughput further: With 16 and 32 operator instances, it is comparable to 8 operator instances. When none of the partial matches can complete (denoted by ``0 cplx''), the throughput scales from 10,400 events/second at 1 operator instance to 108,400 events/second at 16 operator instances (scaling factor 10.4) and 174,300 events/second at 32 operator instances (scaling factor 16.8). 

\textbf{Discussion of the results.} We draw the following conclusions from the results. First of all, our assumptions on the system behavior are backed by the measurements. Further, the different queries impose ``throughput profiles'' that have a similar shape. The scaling behavior in SPECTRE, using the speculation approach, is very different from other event processing systems that have been analyzed in related work. In SPECTRE, the parallelization-to-throughput ratio largely depends on the completion probability of partial matches. This new factor leads to interesting implications when adapting the parallelization degree (i.e., elasticity), which is typically done based on event rates \cite{DeMatteis:2016:KCR:2851141.2851148, 7024105, lohrmann_elastic_2015} or CPU utilization \cite{akidau_millwheel:_2013, CastroFernandez:2013:ISO:2463676.2465282}. Existing elasticity mechanisms do not take into account the completion probability to determine the optimal resource provisioning. Using the described throughput curves, SPECTRE could adapt the number of operator instances based on the current pattern completion probability.

\begin{figure}
\begin{minipage}[t]{0.47\linewidth}
	\begin{overpic}[width=\linewidth]{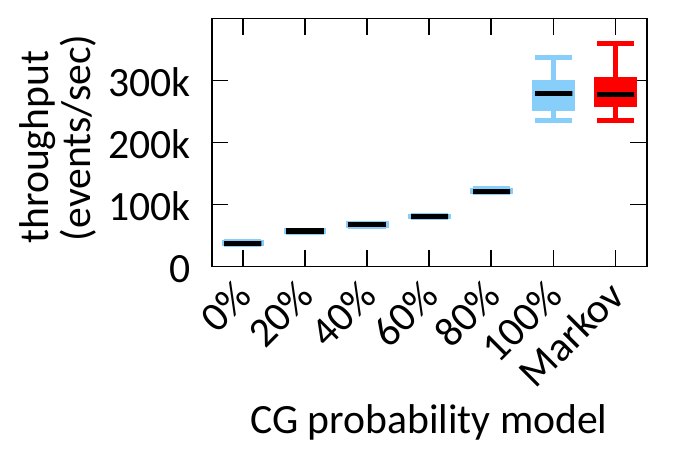}
        		\put(0,0){(a)}
      	\end{overpic}
\end{minipage}%
    \hfill%
\begin{minipage}[t]{0.47\linewidth}
	\begin{overpic}[width=\linewidth]{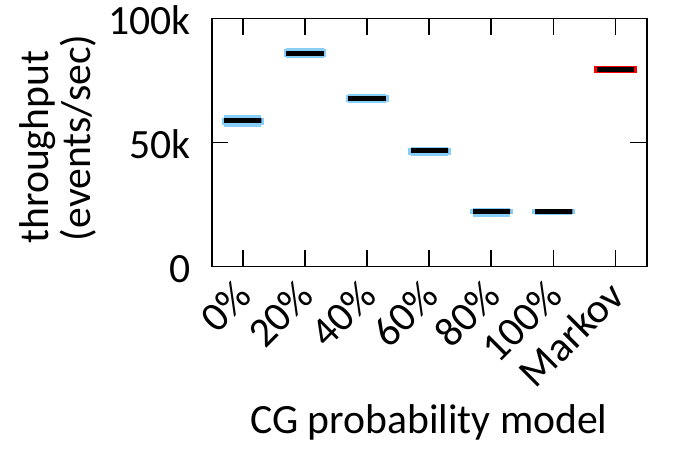}
        		\put(0,0){(b)}
      	\end{overpic}
\end{minipage}%
\hfill
\vspace{-0.1cm}
\caption{Evaluation of Markov Model.}
\label{fig:throughput_markov}
\vspace{-0.3cm}
\end{figure}

\paragraph{Overhead of Speculation}

Here, we analyze the computational and memory overhead of maintaining the dependency tree in the splitter and scheduling the top-$k$ window versions. 

In a first experiment (Q1, NYSE dataset, $q$ = 80, window size = 8,000), we measure how often the splitter can perform a complete cycle of tree maintenance and top-$k$ scheduling per second. The cycle is described as follows: (a) Maintenance: performing all updates on the dependency tree that have been issued since the last maintenance, i.e., creating new consumption groups and window versions and delete dropped ones, and (b) scheduling: schedule the new top-$k$ window versions to the $k$ operator instances according to the updated dependency tree. 

In Figure \ref{fig:performance_worker_threads} (c), the results are depicted. With 1 operator instance, SPECTRE achieves a maintenance and scheduling frequency of 4 million cycles per second. With increasing number of operator instances, the scheduling frequency decreases but is still considerably high, where SPECTRE achieves a scheduling frequency of $650,000$ and $450,000$ times per second with 16 and 32 operator instances, respectively. We conclude that there is some overhead involved in the management of the dependency tree and the scheduling algorithm, but there are no indications that this would become a bottleneck in the system.

Another concern about the dependency tree might be its growth and size in memory. To this end, we measured the maximal number of window versions maintained in the dependency tree at the same time (Q1, NYSE dataset, $q$ = 80, window size = 8,000). The results of the experiments are depicted in Figure \ref{fig:performance_worker_threads} (f). With 1 operator instance, the maximal tree size was at 41 window versions, growing up to 4,332 at 16 operator instances and 6,730 window versions at 32 operator instances. This is not a serious issue in terms of memory consumption. Indeed, the importance of a suitable top-$k$ window version selection becomes obvious here: Determining the $k$ window versions that will survive out of a large number of window versions that will eventually be dropped is a huge challenge, which SPECTRE could handle reasonably well in the performed experiments.

\subsubsection{Markov Model}

After we have discussed the overall system throughput and different factors that impact it, we go into a more detailed analysis of the completion probability model of consumption groups. In particular, we want to know how well the proposed Markov model behaves when the probabilities of complex events are changing. To this end, we perform two different experiments of query Q3 with different ratios of pattern size to window size: A ratio of 0.002 that has a high consumption group completion probability and a ratio of 0.1 that has a lower consumption group completion probability. We employed 32 operator instances and the window size $\mathit{ws}$ was set to 1000 events where a new window is opened every 100 events ($s$ = 100). We compare the proposed Markov model with a probability model that assigns each consumption group a fixed  completion probability. The results of the two experiments are depicted in Figure \ref{fig:throughput_markov} (a) and (b), respectively.

At a ratio 0.002, the completion probability of a consumption group was at $100\%$. Accordingly, assigning a fixed probability of $100\%$ to the consumption groups yielded a throughput of 279,000 events per second, which was significantly better than other fixed probabilities. The Markov model with a throughput of 277,000 events per second proved to be competitive with the best fixed model.

At a ratio of 0.1, the probability of a complex event was at of $32\%$. Accordingly, assigning a fixed probability of $20\%$ to the consumption groups yielded a throughput of 86,000 events per second, which was significantly better than other fixed probabilities. The Markov model with a throughput of 79,000 events per second performed almost as good as the best fixed model. 

From those results, we draw two conclusions. First, the Markov model is able to automatically learn suitable consumption group probabilities in different settings. Second, we can see that wrong probability predictions can cause a large throughput penalty.

\subsubsection{Comparison to T-REX} We have also implemented query Q1 in the T-REX event processing engine \cite{Cugola:2012:CEP:2221990.2222305}. In total numbers, T-REX performed much worse than SPECTRE, reaching a throughput of only about 1,000 events per second. While this shows that the throughput of SPECTRE is competitive, it is worth to mention that both systems are different. T-REX is a general-purpose event processing engine that automatically translates queries into state machines, whereas SPECTRE employs user-defined functions to implement queries which allows for more code optimizations. T-REX does not support event consumptions in parallel processing, while SPECTRE can utilize multi-core machines to scale the throughput. 
\section{Related Work}
\label{sec:Related Work}
In the past decades, a number of different Complex Event Processing systems and languages has been proposed. Besides CEP languages that do not support event consumptions, such as SASE \cite{wu2006high}, the concept of event consumption gained growing importance.
Based on practical use cases, Snoop \cite{Chakravarthy1994} defined 4 different so-called \emph{parameter contexts}, which are predefined combinations of Selection and Consumption Policies. Building on a more systematic analysis of the problem, Zimmer and Unland \cite{754955} proposed an event algebra that differentiated between 5 different Selection and 3 different Consumption Policies that can be combined. Picking up and extending that work, the Amit system \cite{Adi:2004:ASM:988145.988150} allowed for distinct specifications of the Selection and Consumption Policy. Finally, Tesla \cite{Cugola:2010:TFD:1827418.1827427} and its implementation T-REX \cite{Cugola:2012:CEP:2221990.2222305} introduced a formal definition of its supported policies. 
The proposed speculation methods and the SPECTRE framework are applicable to any combination of selection and consumption policies.

The crucial question in exploiting data parallelism in a DCEP operator is how to split the incoming event streams, such that the different partitions, assigned to different operator instances, can be processed in parallel. Besides window-based splitting, as used in SPECTRE, other splitting methods have been proposed. However, they lack the expressiveness to capture temporal relations between events that many DCEP queries expose.

In key-based splitting \cite{Mayer:2016:GRD:2933267.2933509, zeitler2011massive,  Hirzel:2012:PCP:2335484.2335506, CastroFernandez:2013:ISO:2463676.2465282, Gedik2014}, the event stream is split by a key that is encoded in the events, e.g., a stock symbol in algorithmic trading \cite{Hirzel:2012:PCP:2335484.2335506} or a post ID in social network analysis \cite{Mayer:2016:GRD:2933267.2933509}. Different key value ranges are assigned to different operator instances. However, the parallelism is restricted to the number of different key values; moreover, not all pattern definitions exhibit key-based data parallelism. For instance, in example query $Q_E$ (cf. Section \ref{sec:dcep}), events of both stock symbols $A$ and $B$ have to be correlated, so that key-based splitting cannot be applied.

Pane-based splitting has been proposed in stream processing systems \cite{Balkesen2011, Koliousis:2016:SWH:2882903.2882906}. For instance, when the max or median value of a window of 1 minute shall be computed, that window is split into 6 fragments of 10 seconds, the fragments' max or median values are computed in parallel, and the global window's value is aggregated from the fragments' results. This parallel aggregation procedure bases on the idea of pane-based aggregations \cite{Li:2005:NPN:1058150.1058158}. However, DCEP patterns often impose a temporal dependency between the events of a window that hinders the vertical splitting, e.g., when a sequence of events $A$ and $B$ is queried as in example query $Q_E$  (cf. Section \ref{sec:dcep}). Furthermore, additional constraints on the events can be formulated, e.g., $A$ and $B$ have a parameter $x$, such that $A:x > B:x$ (e.g., to detect chart patterns in stock markets \cite{Hirzel:2012:PCP:2335484.2335506}). If the events are scattered among different vertical windows, such dependencies and constraints cannot be analyzed.

Besides data parallelization, \emph{intra-operator parallelization}, also known as pipelining, has been proposed. Internal processing steps that can be run in parallel are identified by deriving operator states and transitions from the query (e.g., state-based approach in \cite{Balkesen:2013:RRI:2488222.2488257}). According to the identified processing steps, the operator logic is split and the processing steps are executed in parallel. This offers only a limited achievable parallelization degree depending on the number of processing steps in the query. For instance, in example query $Q_E$  (cf. Section \ref{sec:dcep}), only 2 processing steps, detecting A and detecting B, are available, leading to a maximum parallelization degree of 2. A common variant of intra-operator parallelization uses \emph{lazy evaluation} techniques on event sequence patterns to increase the operator throughput \cite{Cugola2012205, Kolchinsky:2015:LEM:2675743.2771832}. Those techniques check the event stream for \emph{terminator} events, i.e., the last event of the event sequence in a pattern, and only evaluate preceding events when such a terminator event is found. The underlying assumption is that a terminator event can be determined independently from other events, e.g., solely based on its event type. However, often, sequence patterns depend on the comparison of the events' payload, e.g., a stock quote increasing 3 times in a row; whether a quote is the third in a row that is increasing can only be determined when the two preceding quotes are analyzed. Hence, such techniques are only addressing a subset of possible event patterns. 

Speculation has been widely applied to deal with out-of-order events in stream processing. Mutschler and Philippsen \cite{Mutschler:2014:ASP:2659232.2633686} propose an adaptive buffering mechanism to sort the events before processing them, introducing a \emph{slack time}. When an event arrives outside of the slack time, results are recomputed. However, slack times cannot be used to overcome window dependencies in the event consumption problem: If one window is processed later, all depending windows would also need to be deferred.
Brito et al. \cite{Brito:2008:SOE:1385989.1386023} as well as Wester et al. \cite{Wester:2009:TLR:1558977.1558994} propose transaction-based systems to roll-back processing when out-of-order events arrive. Their systems are not parallel, meaning that they only employ one speculation path for each operator. We also roll-back when window versions reach an inconsistent state. However, we propose a highly parallel multi-path speculation method (not only one path) and employ a probabilistic model to schedule the most promising window versions; hence, our system scales with an increasing number of CPU cores.
Balazinska et al. \cite{balazinska2007moirae} propose a system that quickly emits approximate results that are later refined when out-of-order events arrive. Our model would generally allow to be extended toward supporting probabilistic approximations, as a survival probability is given on the window versions. However, in this paper, we focus on consistent event detection (no false-positives, no false-negatives) and leave approximate applications of our model to the future work.
Brito et al. \cite{5158422} propose for non-deterministic stream processing operators to mark events as speculative before logs have been committed to disc for consistent recovery. The speculative events can be forwarded to successor operators in the operator graph that treat them specifically. In SPECTRE, speculative complex events are kept buffered until the window version is confirmed. We focus on providing deterministic event streams to the successor operators; in particular, we do not assume that subsequent operators or event consumers can handle events that are marked as speculative.

\section{Conclusion}

The SPECTRE system uses window-based data parallelization and optimized speculative execution of interdependent windows to scale the throughput of DCEP operators that impose consumption policies. The novel speculation approach employs a probabilistic consumption model that allows for processing the $k$ most promising window versions by $k$ operator instances in parallel on a multi-core machine. Evaluations of the system show good scalability at a moderate overhead for speculation management. 

\section*{Acknowledgments} This work was funded by DFG grant RO 1086/19-1 (PRECEPT).

\bibliographystyle{ACM-Reference-Format}
\bibliography{multicore} 

\end{document}